\def\arxivbuild{1}
\documentclass{article}
\usepackage{iclr2027_conference,times}

\newif\ifarxiv
\ifdefined\arxivbuild
    \arxivtrue
\else
    \arxivfalse
\fi

\usepackage{amsmath,amsfonts,bm}

\def\eqref#1{equation~\ref{#1}}
\def\1{\bm{1}}

\DeclareMathAlphabet{\mathsfit}{\encodingdefault}{\sfdefault}{m}{sl}
\SetMathAlphabet{\mathsfit}{bold}{\encodingdefault}{\sfdefault}{bx}{n}

\newcommand{\benchmark}{PianoBench}

\newcommand{\dataset}{PianoTalk}

\usepackage{graphicx}
\usepackage{fancyvrb}
\usepackage{fvextra}
\usepackage{booktabs}
\usepackage{capt-of}
\usepackage{wrapfig}
\usepackage{siunitx}
\usepackage{tikz}
\usetikzlibrary{arrows.meta,calc,positioning}
\usepackage{hyperref}
\hypersetup{
    colorlinks=true,
    linkcolor=[rgb]{0,0,0.5},
    filecolor=[rgb]{0,0,0.5},
    urlcolor=[rgb]{0,0,0.5},
    citecolor=[rgb]{0,0,0.5},
    linktocpage=false
}
\usepackage[capitalise,noabbrev,nameinlink]{cleveref}
\usepackage{url}
\usepackage[normalem]{ulem}

\title{PianoBench: Evaluating Musical Perception and Reasoning in Language Models}

\author{Louis Bradshaw\textsuperscript{1}, Irmak Bukey\textsuperscript{2},
\textbf{Daniel Chenyu Lin\textsuperscript{3}, John Thickstun\textsuperscript{3},  Chris Donahue\textsuperscript{2}}\\
\textsuperscript{1}Queen Mary University of London,
\textsuperscript{2}Carnegie Mellon University,
\textsuperscript{3}Cornell University
}

\iclrfinalcopy

\begin{document}

\maketitle
\ifarxiv
    \fancyhead{}
\fi

\begin{abstract}
Multimodal language 
models continue to improve on broad 
music understanding
benchmarks, but which musical features can they perceive and reason about?
We investigate this question by introducing \textit{\benchmark}, a benchmark 
that tasks language models with reasoning about musical performances 
across congruent acoustic and symbolic input representations.
\benchmark{} is derived from our new multimodal, training-scale dataset 
comprising more than 20,000 hours of
interleaved transcripts of music and spoken musical commentary from instructional piano videos.
Open-weight audio-language models achieve at most 52.8\% accuracy on \benchmark{}, and at best only modestly outperform their accuracy on a no-audio control. 
In contrast, 
frontier models using chain of thought achieve up to 93.2\% accuracy when provided transcripts of music as text,
exposing a gap between \emph{perceiving} musical features from audio and
\emph{reasoning} about musical events when they are given explicitly.
We then test whether mid-training on interleaved musical instruction can improve question answering without chain of thought, raising Qwen 3.5 9B Base's zero-shot accuracy on \benchmark{} with music transcripts from 42.8\% to 56.0\% after instruction tuning.
These experiments suggest that instructional videos may represent an overlooked source of multimodal supervision for music.
\ificlrfinal
Available: \url{https://gclef-cmu.org/pianobench}.
\else
Supplement: \url{https://anon834910.github.io/anon-dataset-demo/}.
\fi

\end{abstract}

\section{Introduction}

Multimodal language models (LMs) are increasingly capable of processing inputs beyond text, including images, speech, and general audio \citep{alayrac2022flamingovisuallanguagemodel, kong2024audioflamingonovelaudio, xu2025qwen3omnitechnicalreport, gemmateam2026gemma4technicalreport}.
For music, however, it remains unclear what these systems can perceive from audio versus infer through reasoning over explicit note events.
Open-weight models such as the Audio Flamingo series \citep{kong2024audioflamingonovelaudio, ghosh2025audioflamingo2audiolanguage, ghosh2026audioflamingonextnextgeneration} and Music Flamingo \citep{ghosh2025musicflamingoscalingmusic} perform strongly on broad benchmarks of music understanding from audio \citep{sakshi2024mmaumassivemultitaskaudio, kumar2025mmauprochallengingcomprehensivebenchmark, weck2024muchomusicevaluatingmusicunderstanding}, but much of the supervision behind both training and evaluation originates from track-level annotations (e.g.,~captions, tags).
To what extent these models can understand and articulate localized aspects of music, such as motifs and chord progressions, remains unclear.
Learning to connect these patterns with verbal descriptions and analysis likely requires more closely aligned supervision than track-level captions and tags typically provide.

\begin{figure*}[t]
    \centering
    \includegraphics[width=\textwidth]{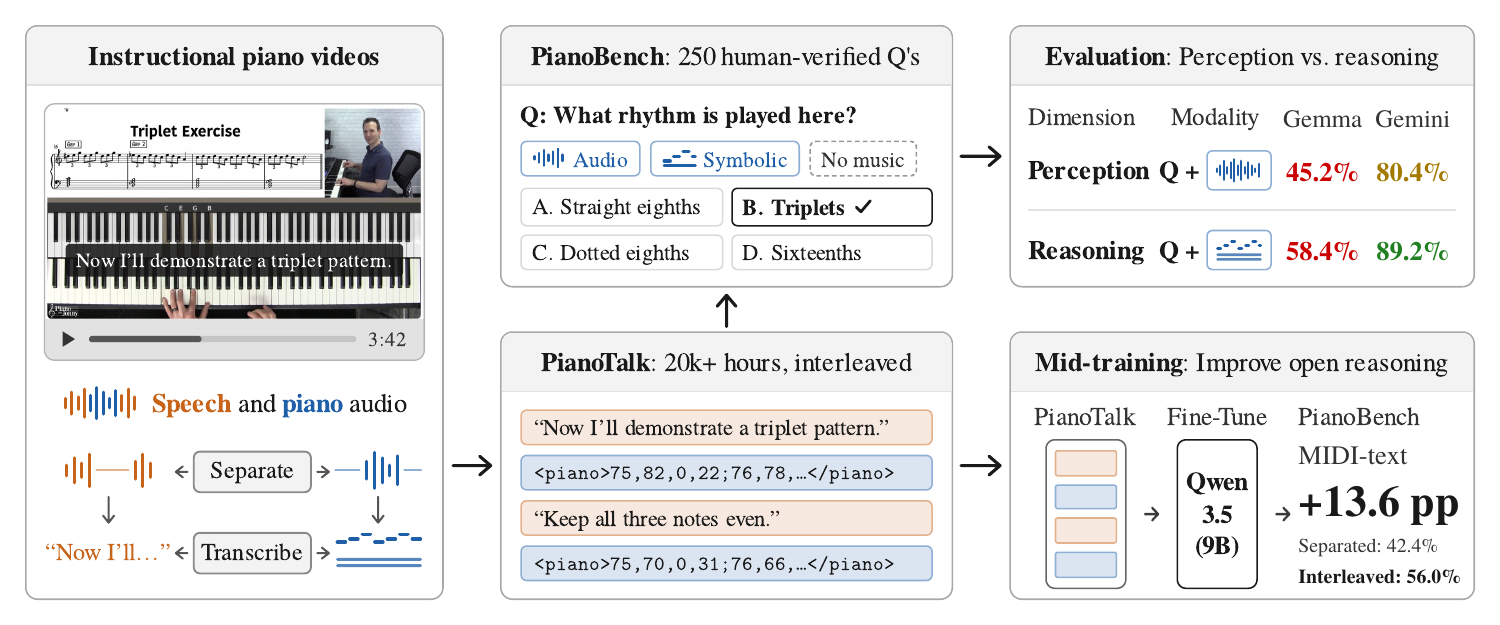}
    \caption{
    We release \benchmark{}, a benchmark for evaluating music perception and reasoning in LLMs. \benchmark{} is derived from \dataset{}, a training-scale (20k hours) dataset of interleaved commentary and music transcripts, which is derived from instructional piano videos. Evaluation results are on \benchmark{} using Gemma 4 12B (thinking) and Gemini 3.1 Pro Preview (medium).
    }
    \label{fig:main}
    \vspace{-3mm}
\end{figure*}

One underexplored setting in which language and music naturally coexist is musical instruction. In a piano tutorial video, a teacher introduces an idea, demonstrates it at the keyboard, and comments on what was just played, so that explanations are adjacent in time to the musical passages they describe. To gather such supervision at scale, we crawl YouTube for instructional piano videos, using an LLM to evaluate their titles and descriptions, then apply source separation \citep{defossez2022hybridspectrogramwaveformsource, solovyev2024benchmarksleaderboardssounddemixing} together with speech and piano transcription \citep{sekoyan2025canary1bv2parakeettdt06bv3efficient, bradshaw2025ariamidi, benetos2018automatic} to convert audio into timestamped commentary text and transcribed music (Figure~\ref{fig:main}). Our resulting dataset, \textbf{\dataset{}}, comprises 138k videos and more than 20,000 hours of interleaved multilingual commentary and music transcripts, in a text format suitable for standard self-supervised LM training.

For evaluation, we construct \textbf{\benchmark{}} from the test split of \dataset{}, comprising 250 human-verified multiple-choice questions (MCQs).
The MCQs in \benchmark{} reference the musical content in short piano passages from 162 videos and range in topic from note and chord details to passage-level ideas such as accompaniment and style. Because every passage has both a recording and a transcript,
we can represent the same musical content as either piano audio, or as a text-based encoding of the symbolic transcript.
The audio condition probes \emph{perception}: whether a model can identify the musical features needed to answer from sound. The text-based transcript condition probes \emph{reasoning}: whether a model can infer those features when the underlying note events are given explicitly. Following \citet{zang2025reallylisteningboostingperceptual}, a no-music control measures the extent to which  answers can be gamed from clues in questions alone.

Evaluating open-weight and proprietary models on \benchmark{} reveals a gap between musical perception from audio and reasoning over explicit musical events. Open-weight audio-language models obtain at most modest gains from audio over a no-music control, with improvements no greater than 6.0 percentage points. By contrast, frontier models using chain of thought (COT) achieve up to 93.2\% accuracy when note events are given explicitly, compared with 80.4\% for the best audio result. These results suggest that models can \emph{reason about} the musical patterns tested by \benchmark{} that remain difficult to \emph{perceive} from audio.

We next test whether interleaved musical instruction improves an open-weight model's zero-shot reasoning over explicit note events without COT. We continue pretraining Qwen 3.5 9B Base on \dataset{}, interleaving chunks of transcribed commentary and a text-compatible representation of the performed piano MIDI in temporal order. A control uses the same chunks in separate commentary-only and music-only documents. Both runs receive general-purpose instruction tuning without music data. Without COT, interleaved training reaches 56.0\% accuracy on \benchmark{}, exceeding instruction tuning alone (42.8\%), separated training (42.4\%), and the official instruction-tuned checkpoint (50.0\%). These results suggest that interleaved musical instruction can develop musical reasoning abilities without COT, providing a complementary route to improving music understanding.

In summary, our contributions are as follows:
\begin{enumerate}
    \item \textbf{Multimodal datasets from musical instruction.} We introduce \dataset{}, a training-scale dataset of interleaved commentary and piano transcripts derived from instructional videos, and \benchmark{}, a human-verified benchmark that presents musical passages as either audio or transcripts.
    \item \textbf{Exposing gaps in music understanding.} Through evaluations on \benchmark{}, we expose a perception gap in open-weight audio-language models, which obtain at most modest gains from audio, and show that frontier models demonstrate strong musical reasoning when given explicit note events and COT.
    \item \textbf{Learning from interleaved instruction.} We show that mid-training on \dataset{} in an interleaved format improves an open model's zero-shot use of note transcripts without COT.
\end{enumerate}

\section{Related Work}
\label{sec:related-work}

Our work most closely relates to families of work around evaluating the musical
understanding capabilities of LMs, as well as work on documented
data recipes for training these capabilities.

\subsection{Benchmarks for Music Understanding}

As music understanding capabilities in language models have progressed, there
has been an increase in benchmarks with questions evaluating music knowledge,
perception, and reasoning abilities. These typically take the form of
multiple-choice questions (MCQs), conditioned on audio or symbolic musical
material such as MIDI and scores.

Among commonly used audio understanding benchmarks, MMAU
\citep{sakshi2024mmaumassivemultitaskaudio}, MMAU-Pro
\citep{kumar2025mmauprochallengingcomprehensivebenchmark}, and MMAR
\citep{ma2025mmarchallengingbenchmarkdeep} all include music questions.
Their questions range from cultural knowledge, mood or style identification,
to more specific questions about rhythm and chord progressions.
MuChoMusic \citep{weck2024muchomusicevaluatingmusicunderstanding} is a
music-only MCQ audio benchmark whose questions were automatically generated
from human-written captions, but verified by humans.
RUListening \citep{zang2025reallylisteningboostingperceptual} identified question
and answer-option priors that models could exploit in MuChoMusic.
It defined a proxy measure based on answer probabilities conditioned on the
question and answer choices alone, without musical input, and proposed a method
for selecting distractors, which inspires our approach to
\benchmark{}. HumMusQA \citep{Weck_2026} is a similar music-only
benchmark that instead has experts generate questions and distractors, and
blindly answer questions to remove weak distractors, without automatic processes.
At a more granular level, MUSE \citep{carone2025musebenchmarkprobingmusic} tests controlled music perception,
MusicTheoryBench
\citep{yuan2024chatmusicianunderstandinggeneratingmusic} uses ABC notation,
and MusICA-MetaBench
\citep{sourada2026musicicareabout} compares performance across audio, ABC notation, and scores, using MusicXML to derive the answers. \benchmark{}
differs by drawing questions from commentary in musical instruction videos,
spanning note and chord details through passage-level ideas such as style.
Matched audio and symbolic inputs let us compare answers based on sound with
answers based on explicit note events.

\subsection{Training Data and Methods for Music Understanding}

Many frontier open-weight language models, e.g., Qwen3-Omni
\citep{xu2025qwen3omnitechnicalreport} and Gemma 4 12B
\citep{gemmateam2026gemma4technicalreport}, report strong performance on
audio-understanding benchmarks alongside general-purpose reasoning
capabilities. However, only a limited amount of information is given regarding
specifics of their training datasets and recipes. Some open-weight models are
more thoroughly documented, including the Audio Flamingo and Nemotron families.
Audio Flamingo (AF) 1, 2, and 3
\citep{kong2024audioflamingonovelaudio,ghosh2025audioflamingo2audiolanguage,goel2025audioflamingo3advancing}
are built on pretrained LLM backbones with separate audio encoders. Their music
data follows a recurring recipe: recordings, captions, metadata, and musical
attributes extracted from audio provide source material, which language models
turn into richer captions and QA pairs. AF3 compiles and releases AudioSkills-XL
using this process, then combines it with other caption and QA datasets for
training.

In follow-up work, Music Flamingo \citep{ghosh2025musicflamingoscalingmusic}
fine-tuned AF3 while expanding this data recipe, creating MF-Skills from
YouTube recordings and publicly releasing its annotations and YouTube IDs.
Music Flamingo also post-trains on MF-Think, a subset of MF-Skills augmented
with generated reasoning traces, followed by reinforcement learning.
Audio Flamingo Next \citep{ghosh2026audioflamingonextnextgeneration} also included
MF-Skills in its music training mix. More recent models use related approaches:
Nemotron 3 Nano Omni 30B-A3B \citep{nvidia2026nemotron3nanoomni} uses a similar
synthetic-data recipe for music QA, while Audex
\citep{kong2026unifiedaudiointelligenceregressing} reports following AF3 and
Audio Flamingo Next in its audio-understanding data.

There has also been work on
adapting pretrained LLMs for
symbolic music understanding. ChatMusician
\citep{yuan2024chatmusicianunderstandinggeneratingmusic} continues training
on MusicPile, which combines ABC-notation with music-related text, including
summaries generated from YouTube track
metadata. MIDI-LLaMA \citep{yang2026midillamainstructionfollowingmultimodalllm}
adds a MIDI encoder and instruction-tunes it on QA and
captions derived from GiantMIDI-Piano
\citep{kong2022giantmidipianolargescalemididataset}. Unlike MF-Skills and other
caption/QA-based audio recipes above, we use the original sequence of
spoken commentary and piano demonstrations from instructional recordings as
text and MIDI, then test continued pretraining on these interleaved sequences.

\clearpage
\section{\dataset{}: Interleaved Piano and Commentary}
\label{sec:dataset}

In this section, we describe the data pipeline used to create \dataset{}. Using this pipeline, we collect interleaved commentary and music transcripts at training scale (20k hours) from publicly available piano instruction videos.
With this paper, we release timestamped speech transcripts and MIDI transcriptions, excluding the source audio.

\begin{figure}[t]
    \centering
    \includegraphics[width=\textwidth]{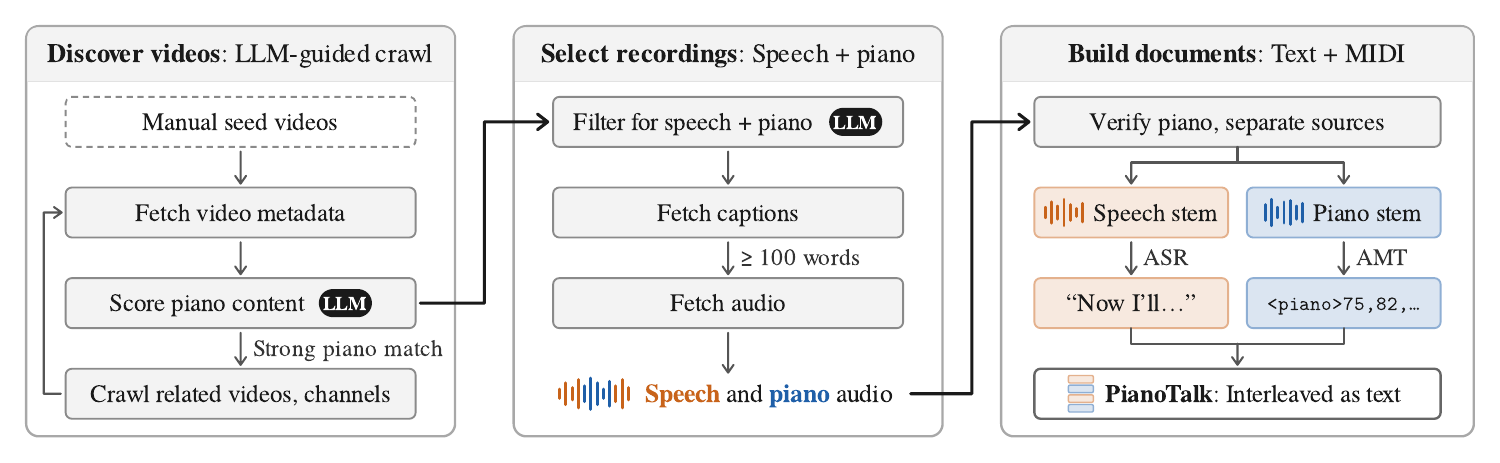}
    \caption{
    \dataset{} data recipe. Starting from seed videos, we crawl related videos
    and YouTube channels. An LLM evaluates video metadata to select videos
    likely to contain both speech and piano.
    Audio from videos with captions is fetched and
    screened for piano content.
    Audio is converted into interleaved speech and music transcripts via source separation, automatic speech recognition (ASR) and automatic music transcription (AMT).
    }
    \label{fig:data-pipeline}
    \vspace{-8pt}
\end{figure}

\subsection{Collecting and Processing Instructional Piano Videos}
\label{sec:pianotalk_collect}

Our data pipeline consists of two stages. First, we identify videos containing
spoken commentary and solo-piano music at scale. We then convert their audio
into time-aligned commentary text and MIDI transcriptions.
\Cref{fig:data-pipeline} provides an overview of the pipeline.

\paragraph{Video discovery.} We search for candidate videos in two stages.
Starting from a small set of manually selected YouTube videos,
an LLM evaluates video titles and descriptions for evidence of piano content,
and we iteratively expand the search by crawling related videos. These
candidates seed a second stage that targets videos containing both spoken
commentary and solo-piano music. In this stage, we crawl related videos and
expand to YouTube channels when at least five of their videos pass the
speech-piano metadata filter or more than 25\% of their scored videos pass the
piano-content filter. We use Gemma 4 26B-A4B
\citep{gemmateam2026gemma4technicalreport} for metadata filtering in this stage.
The metadata scoring criteria, crawl expansion rules, and speech-piano
selection prompt are provided in
\cref{appendix:candidate-discovery,appendix:speech-piano-selection}.

\paragraph{Audio processing.} We discard selected videos whose auto-generated
captions contain fewer than 100 words and download the audio from the remaining
videos. To filter out recordings with insufficient piano content, we use the
piano classifier from \citet{bradshaw2025ariamidi}. The classifier produces a
per-second score trained as a proxy for the wet-dry mix of piano and non-piano
audio. We retain recordings with a mean score of at least 0.3. We then separate
the piano audio using
MVSEP\footnote{\href{https://github.com/ZFTurbo/Music-Source-Separation-Training/releases\#release-v1.0.21}{https://github.com/zfturbo/music-source-separation-training/releases\#release-v1.0.21}}
\citep{solovyev2024benchmarksleaderboardssounddemixing} and transcribe the full
piano stem to MIDI using the transcription model from
\citet{bradshaw2025ariamidi}. Finally, we generate speech transcripts using
Parakeet-TDT-0.6B-v3
\citep{sekoyan2025canary1bv2parakeettdt06bv3efficient} to supplement the
auto-generated captions.

\subsection{MIDI-Text Tokenization}
\label{sec:dataset-tokenizer}

To provide piano input to text-only language models, we encode the transcribed
MIDI as a text representation, which we call \emph{MIDI-text}. This encoding is
designed to be token efficient, information dense, and parseable by existing
LLM text tokenizers. Each MIDI segment is first tokenized using the Aria tokenizer
\citep{bradshaw2025scalingselfsupervisedrepresentationlearning}, which
quantizes note velocities to 12 nonzero levels and onsets and durations to 10\,ms
increments. Each note is then represented by its pitch, velocity, onset, and
duration. A marker is added every 5 seconds, and note onsets are measured from
the most recent marker. This keeps onset values bounded while preserving timing
across the segment. The text header specifies the time units and note fields.
For example, a C-major scale from C4 to C5 with notes 1 second apart and durations
of 150\,ms is represented as follows (colors indicate individual tokens):

\begingroup
\footnotesize
\definecolor{miditoken0}{HTML}{E4D5F5}
\definecolor{miditoken1}{HTML}{D1E8FA}
\definecolor{miditoken2}{HTML}{D5EEDB}
\definecolor{miditoken3}{HTML}{FCE5C8}
\definecolor{miditoken4}{HTML}{F7D4DE}
\setlength{\fboxsep}{0pt}
\newcommand{\mt}[2]{\colorbox{miditoken#1}{\strut #2}}
\noindent
\begin{minipage}[c]{0.37\linewidth}
  \centering
  \includegraphics[width=\linewidth]{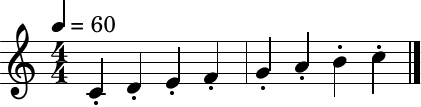}
\end{minipage}%
\hfill$\longrightarrow$\hfill
\begin{minipage}[c]{0.565\linewidth}
\begin{Verbatim}[commandchars=\\\{\}]
\mt{0}{<p}\mt{1}{iano}\mt{2}{>}\mt{3}{unit}\mt{4}{=}\mt{0}{10}\mt{1}{ms}\mt{2}{;}\mt{3}{advance}\mt{4}{=}\mt{0}{5}\mt{1}{s}\mt{2}{;}
\mt{3}{notes}\mt{4}{(m}\mt{0}{idi}\mt{1}{_pitch}\mt{2}{,}\mt{3}{velocity}\mt{4}{,on}\mt{0}{set}\mt{1}{,d}\mt{2}{uration}\mt{3}{):}
\mt{4}{60}\mt{0}{,}\mt{1}{90}\mt{2}{,}\mt{3}{0}\mt{4}{,}\mt{0}{15}\mt{1}{;}\mt{2}{62}\mt{3}{,}\mt{4}{90}\mt{0}{,}\mt{1}{100}\mt{2}{,}\mt{3}{15}\mt{4}{;}\mt{0}{64}\mt{1}{,}\mt{2}{90}\mt{3}{,}\mt{4}{200}\mt{0}{,}\mt{1}{15}\mt{2}{;}
\mt{3}{65}\mt{4}{,}\mt{0}{90}\mt{1}{,}\mt{2}{300}\mt{3}{,}\mt{4}{15}\mt{0}{;}\mt{1}{67}\mt{2}{,}\mt{3}{90}\mt{4}{,}\mt{0}{400}\mt{1}{,}\mt{2}{15}\mt{3}{;}\mt{4}{advance}\mt{0}{;}
\mt{1}{69}\mt{2}{,}\mt{3}{90}\mt{4}{,}\mt{0}{0}\mt{1}{,}\mt{2}{15}\mt{3}{;}\mt{4}{71}\mt{0}{,}\mt{1}{90}\mt{2}{,}\mt{3}{100}\mt{4}{,}\mt{0}{15}\mt{1}{;}\mt{2}{72}\mt{3}{,}\mt{4}{90}\mt{0}{,}\mt{1}{200}\mt{2}{,}\mt{3}{15}\mt{4}{;</}\mt{0}{p}\mt{1}{iano}\mt{2}{>}
\end{Verbatim}
\end{minipage}
\endgroup

\subsection{Dataset Composition and Splits}
\label{sec:dataset-composition}

\begingroup
\setlength{\intextsep}{0pt}
\begin{wrapfigure}{r}{0.48\textwidth}
    \vspace{2pt}
    \centering
    \includegraphics[width=\linewidth]{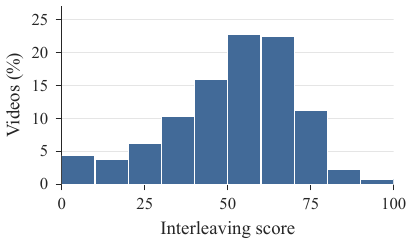}
    \caption{Distribution of interleaving scores across \dataset{} videos,
    measured from chronological speech and MIDI chunks.}
    \label{fig:interleaving-distribution}
\end{wrapfigure}
In total, \dataset{} contains 138,212 videos. The speech intervals span 17,706
hours and the piano intervals span 14,931 hours, including 6,571 hours in
which speech and piano overlap. At least one modality is present for 26,066
hours. The captions cover 44 languages,  English (68.48\%), Spanish (8.41\%),
and French (3.91\%) being the most common caption languages.
\Cref{fig:interleaving-distribution} shows speech-piano interleaving across the
corpus, and the scoring procedure is described in \cref{appendix:interleaving}.
We provide train, validation, and test splits containing
110,554, 13,837, and 13,821 videos, respectively (approximately 80/10/10). All
videos from a given channel are assigned to the same split. 
\par
\WFclear
\endgroup

\section{\benchmark{}: Musical Perception and Reasoning}
\label{sec:benchmark}

\begin{figure*}[t]
    \centering
    \includegraphics[width=\textwidth]{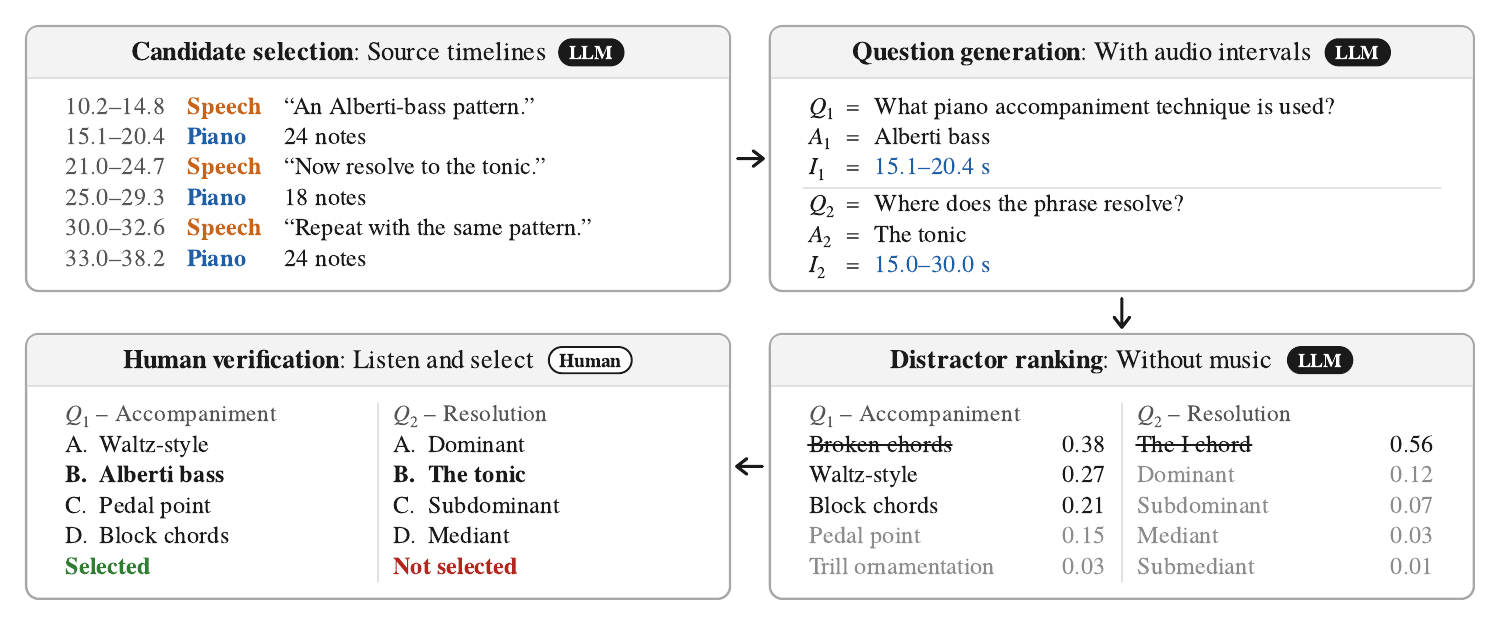}
    \caption{Schematic construction of \benchmark{} (clockwise from top left). Speech and
    piano activity form a source timeline. An LLM with access to the source
    timeline proposes questions, answers,
    audio intervals, and distractors. A model ranks distractors without hearing
    the music. A human checks the passages and answer choices, retaining Q1
    and rejecting Q2.}
    \label{fig:benchmark-pipeline}
\end{figure*}

The instructional videos used to construct \dataset{} provide examples of
solo-piano music with adjacent commentary, but these examples require further
processing before they can be used to evaluate musical capabilities. In this
section, we describe how we process the piano stems, captions, and MIDI
transcriptions from the \dataset{} test split to construct \benchmark{},
a set of 250 human-verified MCQs about short piano passages from 162 videos.
During evaluation, each passage is presented separately as
source-separated piano audio, audio rendered from its MIDI transcription using
a virtual piano (Pianoteq\footnote{We use Pianoteq 8 Stage with the NY Steinway D
Classical Recording preset.}), or
MIDI-text. A no-music control measures what models infer from
the question and choices alone, allowing us to assess the contribution of each
musical representation. Our construction pipeline is shown in
\cref{fig:benchmark-pipeline}.

\textbf{Candidate Selection.} Using the \dataset{} test split, we first process the
MIDI and speech captions into a structured list of contiguous chunks using the
word and note timestamps. We define a chunk as any sequence of events with no
more than 1 second between consecutive notes or 2 seconds between consecutive
words. We format this information into a compressed representation containing
chunk start and end timestamps, speech text, and MIDI-note counts. We
concatenate these representations to produce a coarse, interleaved
timeline for each video. To identify videos with content most relevant for
question generation, we use a lightweight LLM,
Gemma 4 26B-A4B, to rank videos using these timelines. An example
timeline and the ranking prompt are given in
\cref{appendix:benchmark-timelines}.

\textbf{Question Generation.} In order of priority determined by our ranking,
we then use a more powerful LLM, GPT 5.6 Sol, to process timelines into
question-answer pairs with corresponding piano-stem segment timestamps (up to
five per video; see \cref{appendix:question-generation}). As the
language model only has access to coarse indicators of piano activity, it is
forced to rely on the surrounding commentary when
suggesting questions. We follow a procedure inspired by
\citet{zang2025reallylisteningboostingperceptual} to generate distractors.
First, we use GPT 5.6 Terra to generate ten candidate distractors, removing
semantically redundant, potentially correct, and ambiguous choices. For each
candidate distractor, we construct three four-way choice sets containing the correct
answer, that distractor, and two randomly sampled distractors from the remaining
pool. Without access to the audio or MIDI, we record the probability that
Gemma 4 12B assigns to selecting that distractor as the correct answer in each
choice set. We define the distractor's adversarial score as the mean of this
probability across the three sets. The prompts for generation and scoring are
given in \cref{appendix:distractor-generation}.

\textbf{Human Verification.} To ensure accuracy and improve the diversity of
the question-answer pairs, we manually inspect and verify the candidate
questions\footnote{The human annotator was an author with prior musical
training.}, listening to both the piano stem and MIDI rendering. We aimed to
select a diverse set of questions with one unambiguous answer for all input
representations, rejecting candidate questions that are incorrect,
unverifiable, or of low quality. We preferred higher-scoring distractors while
removing duplicate, irrelevant, or misleading choices missed by the distractor
generator. An example benchmark question is shown in
\cref{appendix:benchmark-example}.

The final music clips are 2--20 seconds long, with a median of 11 seconds.
During human verification, we favored questions with clear piano stems and
reliable MIDI transcriptions. We nevertheless identified a few faulty
transcriptions after using \benchmark{} to benchmark models. These are tagged in the released
dataset, but were included in the results reported in
\cref{sec:benchmark-results}. We discuss their effect on evaluation in
\cref{sec:benchmarking-discussion}.

\section{Diagnosing Musical Understanding in Current Models}
\label{sec:benchmarking-models}

We use \benchmark{} to evaluate the music understanding capabilities of
audio-language models and text-only language models. We compare performance
with source-separated piano audio, Pianoteq rendering, and MIDI-text, as well
as with a no-music control. The control measures
how well models can answer from clues in the questions and answer choices
alone. Accuracy improvements over this control (denoted by $\Delta$) measure
the benefit of providing each musical input. Comparing MIDI-text with audio
reveals the gap between reasoning over explicit note events and identifying
the musical features needed to answer from sound. Comparing the
source-separated piano stem with its Pianoteq rendering tests whether audio
performance is consistent across the two representations.

\subsection{Experimental Setup}
\label{sec:experimental-setup}

We evaluate three broad model classes: audio-language models, specialist audio
and music-understanding models, and large language models with
COT reasoning capabilities. We include Nemotron, Audex, and the Audio Flamingo
and Music Flamingo models, whose publicly documented training recipes use
music captions, question-answer pairs, and automatic
speech recognition data. We used model-specific inference configurations,
following author-recommended settings where provided. For configurations
without a thinking mode for chain-of-thought reasoning, we generally used
greedy decoding for MCQ answering.
We report the exact hyperparameters and inference route for each model in
\cref{appendix:model-inference}. During benchmark construction, we shuffled
the answer choices once per question and held their order fixed across models
and input conditions. To check that models could receive and process the
audio input, we asked each audio model to identify the sound in several
held-out piano clips and confirmed that it identified piano music. During
benchmarking, we requested responses in JSON format and extracted the answer.
To accommodate instruction-following failures observed in some models, we
also accepted an isolated valid answer letter. If neither format could be
parsed, we resubmitted the prompt for up to five attempts in total. We report
accuracy under each input condition and the
percentage-point change ($\Delta$) from the corresponding no-music accuracy.
The prompts for all input conditions are provided in
\cref{appendix:evaluation-prompts}.

\subsection{Results}
\label{sec:benchmark-results}

\begin{table*}[t]
\centering
\renewcommand{\arraystretch}{1.25}
\setlength{\tabcolsep}{4pt}
\fontsize{9.7}{8}\selectfont
\sisetup{detect-all, table-number-alignment=center}
\newcommand{\audioonlyflag}{\smash{$^{\dagger}$}}
\newcommand{\errorflag}{\smash{\rlap{$^{\ddagger}$}}}
\newcommand{\thinkingflag}{\smash{$^{\ast}$}}
\begin{tabular*}{\textwidth}{@{\extracolsep{\fill}}l
 S[table-format=2.1]
 S[table-format=2.1] >{\scriptsize}S[table-format=+2.1, retain-explicit-plus]
 S[table-format=2.1] >{\scriptsize}S[table-format=+2.1, retain-explicit-plus]
 S[table-format=2.1] >{\scriptsize}S[table-format=+2.1, retain-explicit-plus]@{}}
\toprule
\textbf{Model} & {\textbf{No music}} & \multicolumn{2}{c}{\textbf{MIDI-text}} & \multicolumn{2}{c}{\textbf{Piano synth}} & \multicolumn{2}{c}{\textbf{Piano stem}} \\
\cmidrule(lr){2-2} \cmidrule(lr){3-4} \cmidrule(lr){5-6} \cmidrule(lr){7-8}
& {Acc.}
& {Acc.} & {\fontsize{9.7}{8}\selectfont $\Delta$}
& {Acc.} & {\fontsize{9.7}{8}\selectfont $\Delta$}
& {Acc.} & {\fontsize{9.7}{8}\selectfont $\Delta$} \\
\midrule
\multicolumn{8}{@{}l}{\textit{Open-weight audio-language models}} \\
\addlinespace[0.5ex]
Qwen 3 Omni 30B-A3B (thinking) & 42.4 & 47.6 & +5.2 & 48.4 & +6.0 & 46.8 & +4.4 \\
Qwen 3 Omni 30B-A3B (non-thinking) & 44.8 & 47.2 & +2.4 & 50.8 & +6.0 & 49.2 & +4.4 \\
Gemma 4 12B (thinking) & 44.8 & 58.4 & +13.6 & 46.4 & +1.6 & 45.2 & +0.4 \\
Gemma 4 12B (non-thinking)\thinkingflag & {-} & {-} & {-} & {$\underline{51.2}$} & {-} & {$\mathbf{51.6}$} & {-} \\
Nemotron 3 Nano Omni 30B-A3B & 46.4 & 45.6 & -0.8 & 47.2 & +0.8 & 48.4 & +2.0 \\
Nemotron Audex 30B-A3B & 41.2 & 48.8 & +7.6 & 42.8 & +1.6 & 39.6 & -1.6 \\
Audio Flamingo 3\audioonlyflag & {-} & {-} & {-} & 38.8 & {-} & 37.2 & {-} \\
Music Flamingo\audioonlyflag & {-} & {-} & {-} & 45.6 & {-} & 42.4 & {-} \\
Inkling$_{\text{small}}$ (medium) & {$\mathbf{51.6}$} & {$\mathbf{68.8}$} & +17.2 & {$\mathbf{52.8}$} & +1.2 & {$\underline{50.4}$} & -1.2 \\
Inkling (medium) & {$\underline{47.2}$} & {$\underline{66.4}$} & +19.2 & 47.6 & +0.4 & 46.8 & -0.4 \\
\midrule
\multicolumn{8}{@{}l}{\textit{Proprietary audio-language models}} \\
\addlinespace[0.5ex]
Gemini 3.1 Pro Preview (medium) & {$\mathbf{63.6}$} & {$\underline{89.2}$} & +25.6 & {$\mathbf{78.4}$} & +14.8 & {$\mathbf{80.4}$} & +16.8 \\
Qwen 3.5 Omni Plus & 52.0 & 57.2 & +5.2 & 62.4 & +10.4 & 63.2 & +11.2 \\
Muse Spark 1.2 (medium) & 48.8 & 86.8 & +38.0 & 56.8 & +8.0 & 55.6 & +6.8 \\
Gemini 3.7 Flash (medium) & {$\underline{62.0}$} & {$\mathbf{90.0}$} & +28.0 & 63.6 & +1.6 & {$\underline{66.8}$} & +4.8 \\
Gemini 3.8 Flash (medium) & {$\underline{62.0}$} & 88.8 & +26.8 & {$\underline{66.4}$} & +4.4 & 64.0 & +2.0 \\
\midrule
\multicolumn{8}{@{}l}{\textit{Language models evaluated on MIDI only}} \\
\addlinespace[0.5ex]
Qwen 3.5 9B (thinking) & 43.6 & 70.0 & +26.4 & {-} & {-} & {-} & {-} \\
Qwen 3.5 9B (non-thinking) & 46.4 & 50.0 & +3.6 & {-} & {-} & {-} & {-} \\
GPT 5.6 Terra (high) & 47.2 & 85.2 & +38.0 & {-} & {-} & {-} & {-} \\
GPT 5.6 Sol (none) & 52.4 & 70.4 & +18.0 & {-} & {-} & {-} & {-} \\
GPT 5.6 Sol (medium) & 56.0 & 85.6 & +29.6 & {-} & {-} & {-} & {-} \\
GPT 5.6 Sol (high) & 56.4 & 87.2 & +30.8 & {-} & {-} & {-} & {-} \\
GPT 5.6 Sol (xhigh) & {$\mathbf{58.8}$} & {$\underline{88.0}$} & +29.2 & {-} & {-} & {-} & {-} \\
GPT 6 Astra (high) & {$\underline{56.8}$} & {$\mathbf{93.2}$} & +36.4 & {-} & {-} & {-} & {-} \\
\bottomrule
\end{tabular*}
\caption{Zero-shot accuracy (\%) on the 250 \benchmark{} MCQs. Models are
evaluated with the question and choices alone (no music), MIDI-text,
Pianoteq-synthesized audio, and the source-separated piano stem, where
supported. $\Delta$ denotes the
percentage-point change from that model's no-music accuracy. $\dagger$: Audio
Flamingo 3 and Music Flamingo had inconsistent instruction following on
text-only inputs. $\ast$: Gemma 4 12B continued to produce chain-of-thought on
MIDI-text and no-music inputs despite thinking being disabled; we therefore
omit these results. \textbf{Bold} is best per column / section, \underline{underline} is second best.}
\label{tab:benchmark-results}
\end{table*}

\paragraph{Audio input.} \Cref{tab:benchmark-results} reports accuracy for the
evaluated models across input conditions. Improvements over the no-music
control differ substantially between open-weight and proprietary audio-language
models. Among open-weight models with a no-music control, audio changes
accuracy by \textminus1.6 to +6.0 percentage points (pp). All proprietary
audio-language models improve over the control in both audio conditions.
Gemini 3.1 Pro achieves the highest audio accuracy, at 80.4\%
($\Delta$ = +16.8 pp), on source-separated piano audio. Qwen 3.5 Omni Plus
\citep{qwenteam2026qwen35omnitechnicalreport} achieves the next-largest improvement in this condition ($\Delta$ = +11.2 pp).
Without no-music results for comparison, Audio Flamingo achieves 37.2\%
piano-stem accuracy and Music Flamingo achieves 42.4\%. Accuracy is
broadly similar between the Pianoteq and source-separated piano audio
conditions across models.

\paragraph{MIDI-text input.} MIDI-text produces the highest accuracy for eight
of the eleven audio-language models that benefit from musical input (a gain of
at least 5 pp over the no-music control in any input condition). The exceptions
are Qwen 3 Omni in both thinking and non-thinking configurations, and Qwen 3.5
Omni Plus. COT reasoning further improves accuracy on MIDI-text. GPT 5.6 Sol
achieves 70.4\% without thinking, rising to 85.6\%, 87.2\%, and 88.0\% at medium,
high, and extra-high reasoning effort, respectively. GPT 6 Astra with high
reasoning effort achieves the highest overall accuracy, at 93.2\%
($\Delta$ = +36.4 pp).

\paragraph{No-music control.}
Accuracy without musical input ranges from 41.2\% (Nemotron Audex) to
63.6\% (Gemini 3.1 Pro), well above the 25\% chance level. These results show
that models can select correct answers significantly above-chance rates using only the
questions and answer choices, despite our efforts to generate robust
distractors. Such cues may have been introduced during human distractor
selection. Models with strong general reasoning and multimodal capabilities,
such as the Gemini 3 family, achieve the highest control accuracies.
GPT 5.6 Terra generated the distractors and Gemma 4 12B ranked them.
Both models' control accuracies fall within the range of other models in their groups.
Because control accuracy varies across models, we assess the benefit of each
musical input using its improvement over the same model's no-music accuracy
($\Delta$), rather than absolute accuracies.

\subsection{Discussion}
\label{sec:benchmarking-discussion}

Taken together, the results reveal differences between models' ability to reason
about explicit musical events and their ability to identify those events from
audio. The limited audio performance of Audex, Audio Flamingo, and Music
Flamingo suggests that scaling current open training recipes based on music
captions and question-answer pairs may be insufficient to develop the musical
understanding capabilities tested by \benchmark{}. The stronger audio gains of
Gemini 3.1 Pro and Qwen 3.5 Omni show that substantially better performance is
possible. Unlike the openly documented recipes, however, their training
mixtures are not publicly described in comparable detail. Additional training
ingredients that more closely connect audio with descriptions of localized
musical events may therefore be needed to reproduce frontier audio performance
in open models.

When note events are given explicitly as MIDI-text, models using COT achieve
strong musical reasoning performance, reaching up to 93\% accuracy on
\benchmark{}. Explicit note events alone are not sufficient, however.
Qwen 3.5 9B and GPT 5.6 Sol both perform substantially better with COT than
without it, suggesting that reasoning helps connect note events to higher-level
musical concepts. However, enabling thinking in Qwen 3 Omni and Gemma 4 12B
does not improve audio accuracy in either condition, suggesting that COT alone
does not overcome the difficulty of recovering musical features from sound.
In the next section, we investigate whether training on interleaved musical
instruction can help models learn connections between note events and musical
concepts, improving MIDI-text performance without COT.

We also examined the 17 questions that GPT 6 Astra answered incorrectly.
Four involved transcription errors missed during human verification, and one
had an incorrect human-selected reference answer\footnote{These
questions remain included in the reported accuracy.}. The remaining errors
tended to involve higher-level musical ideas across passages containing many
notes, such as Q: “What type of meter or dance-like accompaniment
underpins this passage?”; A: “A waltz-like accompaniment in triple meter, with
a triplet feel.” We list these questions in \cref{appendix:astra-errors}.
This failure analysis suggests that interpreting higher-level musical structure
remains challenging even when note events are given explicitly and COT is available.

\clearpage
\section{Learning from Interleaved Musical Instruction}
\label{sec:mid-training}

Our experiments with \benchmark{} reveal that explicit note events support
strong performance in models using COT, while models without COT make less use
of MIDI-text. In this section, we ask to what degree \dataset{} can help models
learn to reason over this representation without COT. One approach would be to
convert the paired commentary text and MIDI transcriptions into automatically
generated supervised descriptions or question-answer pairs, like the
music-understanding data recipes discussed previously. We instead adopt a
continued-pretraining approach, using the interleaved commentary and MIDI-text
as self-supervised training data.

We test whether placing temporally aligned speech and MIDI chunks within
interleaved documents improves downstream reasoning over MIDI-text after
instruction tuning, compared with training on the same chunks as separate
speech and MIDI documents. Focusing on MIDI-text keeps the model architecture
and training objective fixed. This allows us to investigate interleaving
without introducing an audio encoder or adapting pretrained audio models and
their training recipes.

\subsection{Mid-Training and Instruction Tuning}

\begingroup
\setlength{\intextsep}{0pt}
\begin{wrapfigure}{r}{0.48\textwidth}
    \vspace{5pt}
    \centering
    \includegraphics[width=\linewidth]{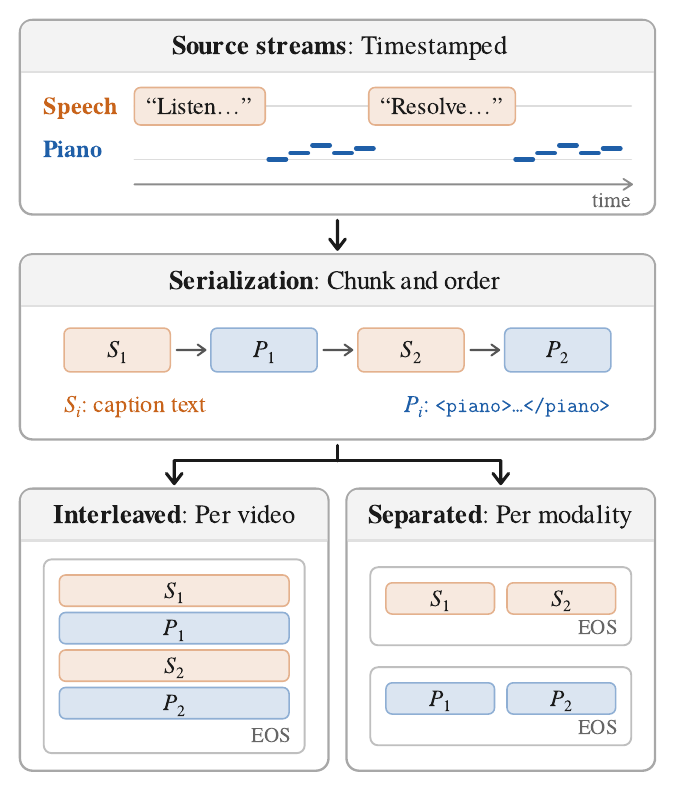}
    \caption{Construction of mid-training documents. Interleaved and
    separated training setups use the same speech and piano chunks, organized
    either chronologically in one interleaved document or as separate
    speech-only and piano-only documents.}
    \label{fig:mid-training-pipeline}
\end{wrapfigure}
We mid-train (i.e., continue pretraining) Qwen 3.5 9B Base on the \dataset{}
training split using three training data setups: speech only, separated speech
and MIDI-text, and interleaved speech and MIDI-text. Comparing the interleaved
and separated setups tests the benefit of preserving the temporal relationship
between commentary and musical events. The separated setup provides exposure
to MIDI-text, but without the accompanying commentary in the same document.
The speech-only setup controls for gains from instructional speech alone.

For mid-training, we split the speech and MIDI streams at gaps longer than
2 seconds. Within longer passages, we add candidate boundaries at word or note
onsets at least 5 seconds apart. We then combine the chunks in chronological order, merging
consecutive chunks of the same modality. This preserves continuous
single-modality passages while producing finer-grained interleaving where
speech and piano overlap. We concatenate the chunks as newline-separated text.
We form the separated documents by splitting each final interleaved document
into speech-only and MIDI-only lines, without additional segmentation or
merging. Both training setups therefore use the same chunks.
\Cref{fig:mid-training-pipeline} visualizes this process; the full construction
and packing procedure is given in \cref{appendix:mid-training-documents}.

\par
\WFclear
\endgroup

We apply the same standard instruction-tuning phase to each
checkpoint, finetuning on a mixture of general instruction-following data (40k
sampled from the Tulu 3 SFT Mixture; \citealp{lambert2025tulu3pushingfrontiers})
and four-way multiple-choice questions (10k sampled from MMLU
\texttt{auxiliary\_train} split;
\citealp{hendrycks2021measuringmassivemultitasklanguage}), reformatted to the
response format of \benchmark{}. We chose Qwen 3.5 9B because it scores only
50.0\% ($\Delta$ = +3.6 pp) with MIDI-text without COT, despite reaching
70.0\% ($\Delta$ = +26.4 pp) with thinking enabled. We also retain the original base checkpoint as an SFT-only
baseline. During instruction tuning we include neither MIDI-text
question-answering data nor chain-of-thought supervision, allowing us to assess
to what degree mid-training improves a smaller model's zero-shot musical
question answering on \benchmark{} without COT.

\subsection{Experiments}

\paragraph{Experimental Setup.} For each training data setup, we mid-trained
Qwen 3.5 9B Base using the AdamW optimizer for one epoch over the corresponding
document set. We used a learning rate of 2e\textminus5 with a
5\% warmup, followed by cosine decay to 10\% of the peak learning rate. We used a sequence length of
4,096 tokens with an effective batch size of 2,621,440 tokens per gradient
update. In total the speech-only, separated, and interleaved training corpora
contained 169 million, 2,988 million, and 2,987 million tokens, respectively.
In the instruction-tuning stage, we trained for one epoch over 50,000 examples
with an effective batch size of 128 examples per update. We used the same peak
learning rate of 2e\textminus5, 5\% warmup, and cosine decay to 10\% of the peak
learning rate.

\begingroup
\setlength{\intextsep}{0pt}
\stepcounter{footnote}
\begin{wraptable}{r}{0.5\textwidth}
    \centering
    \begingroup
    \renewcommand{\arraystretch}{1.25}
    \setlength{\tabcolsep}{4pt}
    \fontsize{9.7}{8}\selectfont
    \sisetup{detect-all, table-number-alignment=center}
    \begin{tabular*}{\linewidth}{@{\extracolsep{\fill}}l
        S[table-format=2.1]
        S[table-format=2.1]
        >{\scriptsize}S[table-format=+2.1, retain-explicit-plus]@{}}
        \toprule
        \textbf{Condition} & {\textbf{No music}} & \multicolumn{2}{c}{\textbf{MIDI-text}} \\
        \cmidrule(lr){3-4}
        & {Acc.} & {Acc.} & {\fontsize{9.7}{8}\selectfont $\Delta$} \\
        \midrule
        Official Instruct & 46.4 & 50.0 & +3.6 \\
        Official Instruct + COT & 43.6 & 70.0 & +26.4 \\
        Base-SFT & 40.8 & 42.8 & +2.0 \\
        \midrule
        Speech only & 37.2 & 37.6 & +0.4 \\
        Separated & 42.4 & 42.4 & +0.0 \\
        Interleaved & 45.6 & \bfseries 56.0 & \bfseries +10.4 \\
        \bottomrule
    \end{tabular*}
    \par
    \endgroup
    \caption{\benchmark{} accuracy (\%) after mid-training and instruction
    tuning of Qwen 3.5 9B. $\Delta$ is the gain over no music (pp). Official
    Instruct is \texttt{Qwen/Qwen3.5-9B}\protect\footnotemark[\value{footnote}]; Base-SFT is its base variant after
    our SFT. Responses use greedy decoding without chain of thought, except
    for Official Instruct + COT.}
    \label{tab:mid-training-results}
\end{wraptable}
\footnotetext{\href{https://huggingface.co/Qwen/Qwen3.5-9B}{https://huggingface.co/Qwen/Qwen3.5-9B}}

\paragraph{Results.} \Cref{tab:mid-training-results} summarizes performance after mid-training and
instruction tuning. Interleaved training achieves 56.0\% accuracy with
MIDI-text ($\Delta$ = +10.4 pp), compared
with gains of 0.0 pp for separated training and 0.4 pp for speech-only
training. This provides evidence that preserving the temporal relationship
between instructional speech and musical events improves downstream use of
MIDI-text beyond exposure to the same content in separate documents.

Our SFT-only baseline achieves 42.8\% accuracy with MIDI-text, compared with
50.0\% for the official instruction-tuned checkpoint. Despite this lower
baseline, interleaved mid-training followed by our SFT exceeds the official
checkpoint without COT by 6.0 pp. With COT, however, the official checkpoint
reaches 70.0\% ($\Delta$ = +26.4 pp). These results suggest that training on
interleaved musical instruction and improving COT reasoning are promising
parallel axes for improving musical question answering.
\par
\WFclear
\endgroup

\section{Conclusion}
\label{sec:conclusion}

In this paper, we have introduced \dataset{} and \benchmark{}, a dataset and
benchmark for learning and evaluating musical understanding across audio and
symbolic inputs. Our evaluations show that strong performance on broad music
benchmarks does not necessarily extend to localized musical tasks: open-weight
audio-language models obtain only modest gains from audio on \benchmark{}.
For models with publicly documented training recipes, this suggests that
existing supervision may be insufficient to develop these capabilities.
Self-supervised training on interleaved music and language, available at scale
through pipelines such as \dataset{}'s, may help address this limitation.

The other main takeaway is that frontier language models perform remarkably
well in a zero-shot setting when given explicit note events as MIDI-text,
connecting individual notes to higher-level musical concepts through
general-purpose reasoning. This
suggests that these models could help generate fine-grained audio supervision,
for example by using transcriptions to construct question-answer pairs about
the corresponding recordings. Our mid-training experiments also show that
interleaved commentary and MIDI-text improve musical question answering
without COT, beyond training on the same content separately. Learning from
such supervision and improving general COT reasoning therefore appear to offer
complementary directions for improving musical understanding. Extending this
supervision to audio could help address the perception limitations identified
here, while scaling both components together may support broader MIR
capabilities.

We hope that our dataset, benchmark, and findings support further work on
connecting musical perception with the reasoning capabilities already
demonstrated by language models.

\clearpage
\ificlrfinal
\section*{Acknowledgements}

This work was supported by the Humanities and AI Virtual Institute (HAVI) at
Schmidt Sciences and UKRI/EPSRC under grant EP/S022694/1. We thank the CMU
Foundation and Language Model (FLAME) Center and HAVI for providing the
computing resources used in our mid-training experiments. Irmak Bukey is
supported by the Waibel Presidential PhD Fellowship. We also thank Simon Colton
for his advice and supervision of Louis.
\fi

\section*{AI Use Statement}

In this work, we use generative AI tools for writing and editing code used in
our dataset, benchmark, and experiments, and for generating and iterating on
figures in the paper. We have also used LLMs in our dataset and benchmark
construction pipelines (as described in \cref{sec:pianotalk_collect} and
\cref{sec:benchmark}). Specifically, we used LLM-based agents to develop and
refine prompts. We used LLMs as grammar checkers to assist with paper writing.
The authors wrote the main text, with LLM assistance for language editing. LLMs
were used more extensively to assist with writing and formatting the
appendices. We have reviewed all AI-assisted work. We take responsibility for
the final content of this work, including text, claims or artifacts produced
with the aid of generative AI.

\section*{Ethics Statement}

Our work sits within the broader landscape of music and AI, and while there are
some ethical considerations worth noting, we believe that the associated
ethical risks are low. Our dataset is built from publicly available internet
videos, which is consistent with common practice in large-scale data collection
efforts in the field. We do not release the audio from these videos. We instead
release a transformed version of the audio, the text and MIDI data we extract.
Our focus is on evaluating and improving music understanding models (where
music is the \emph{input}) as a step towards supporting musicians in their
creative process, as opposed to training generative models (where music is the
\emph{output}) on copyrighted content. The latter raises far more significant
concerns regarding the replacement of human creativity and labor. Furthermore,
while our dataset covers 44 languages, it predominantly features widely spoken
Western languages. This may introduce geographic and cultural bias.

\bibliography{references}

@misc{kong2024audioflamingonovelaudio,
      title={{Audio Flamingo}: A Novel Audio Language Model with Few-Shot Learning and Dialogue Abilities},
      author={Zhifeng Kong and Arushi Goel and Rohan Badlani and Wei Ping and Rafael Valle and Bryan Catanzaro},
      year={2024},
      eprint={2402.01831},
      archivePrefix={arXiv},
      primaryClass={cs.SD},
      url={https://arxiv.org/abs/2402.01831}, 
}

@misc{alayrac2022flamingovisuallanguagemodel,
      title={Flamingo: a Visual Language Model for Few-Shot Learning}, 
      author={Jean-Baptiste Alayrac and Jeff Donahue and Pauline Luc and Antoine Miech and Iain Barr and Yana Hasson and Karel Lenc and Arthur Mensch and Katie Millican and Malcolm Reynolds and Roman Ring and Eliza Rutherford and Serkan Cabi and Tengda Han and Zhitao Gong and Sina Samangooei and Marianne Monteiro and Jacob Menick and Sebastian Borgeaud and Andrew Brock and Aida Nematzadeh and Sahand Sharifzadeh and Mikolaj Binkowski and Ricardo Barreira and Oriol Vinyals and Andrew Zisserman and Karen Simonyan},
      year={2022},
      eprint={2204.14198},
      archivePrefix={arXiv},
      primaryClass={cs.CV},
      url={https://arxiv.org/abs/2204.14198}, 
}

@misc{qwenteam2026qwen35omnitechnicalreport,
      title={{Qwen3.5-Omni} Technical Report},
      author={{Qwen Team}},
      year={2026},
      eprint={2604.15804},
      archivePrefix={arXiv},
      primaryClass={cs.CL},
      url={https://arxiv.org/abs/2604.15804}, 
}

@misc{ghosh2025audioflamingo2audiolanguage,
      title={{Audio Flamingo 2}: An Audio-Language Model with Long-Audio Understanding and Expert Reasoning Abilities},
      author={Sreyan Ghosh and Zhifeng Kong and Sonal Kumar and S Sakshi and Jaehyeon Kim and Wei Ping and Rafael Valle and Dinesh Manocha and Bryan Catanzaro},
      year={2025},
      eprint={2503.03983},
      archivePrefix={arXiv},
      primaryClass={cs.SD},
      url={https://arxiv.org/abs/2503.03983}, 
}

@misc{ghosh2025musicflamingoscalingmusic,
      title={{Music Flamingo}: Scaling Music Understanding in Audio Language Models},
      author={Sreyan Ghosh and Arushi Goel and Lasha Koroshinadze and Lee, {Sang-gil} and Zhifeng Kong and Joao Felipe Santos and Ramani Duraiswami and Dinesh Manocha and Wei Ping and Mohammad Shoeybi and Bryan Catanzaro},
      year={2025},
      eprint={2511.10289},
      archivePrefix={arXiv},
      primaryClass={eess.AS},
      url={https://arxiv.org/abs/2511.10289}, 
}

@misc{sakshi2024mmaumassivemultitaskaudio,
      title={{MMAU}: A Massive Multi-Task Audio Understanding and Reasoning Benchmark},
      author={S Sakshi and Utkarsh Tyagi and Sonal Kumar and Ashish Seth and Ramaneswaran Selvakumar and Oriol Nieto and Ramani Duraiswami and Sreyan Ghosh and Dinesh Manocha},
      year={2024},
      eprint={2410.19168},
      archivePrefix={arXiv},
      primaryClass={eess.AS},
      url={https://arxiv.org/abs/2410.19168}, 
}

@misc{weck2024muchomusicevaluatingmusicunderstanding,
      title={{MuChoMusic}: Evaluating Music Understanding in Multimodal Audio-Language Models},
      author={Benno Weck and Ilaria Manco and Emmanouil Benetos and Elio Quinton and George Fazekas and Dmitry Bogdanov},
      year={2024},
      eprint={2408.01337},
      archivePrefix={arXiv},
      primaryClass={cs.SD},
      url={https://arxiv.org/abs/2408.01337}, 
}

@misc{zang2025reallylisteningboostingperceptual,
      title={Are you really listening? {Boosting} Perceptual Awareness in {Music-QA} Benchmarks},
      author={Yongyi Zang and Sean O'Brien and Taylor Berg-Kirkpatrick and Julian McAuley and Zachary Novack},
      year={2025},
      eprint={2504.00369},
      archivePrefix={arXiv},
      primaryClass={cs.SD},
      url={https://arxiv.org/abs/2504.00369}, 
}

@misc{bradshaw2025ariamidi,
      title={{Aria-MIDI}: A Dataset of Piano {MIDI} Files for Symbolic Music Modeling},
      author={Louis Bradshaw and Simon Colton},
      year={2025},
      eprint={2504.15071},
      archivePrefix={arXiv},
      primaryClass={cs.SD},
      url={https://arxiv.org/abs/2504.15071}, 
}

@misc{lambert2025tulu3pushingfrontiers,
      title={{Tulu 3}: Pushing Frontiers in Open Language Model Post-Training},
      author={Nathan Lambert and Jacob Morrison and Valentina Pyatkin and Shengyi Huang and Hamish Ivison and Faeze Brahman and Lester James V. Miranda and Alisa Liu and Nouha Dziri and Shane Lyu and Yuling Gu and Saumya Malik and Victoria Graf and Jena D. Hwang and Jiangjiang Yang and Ronan Le Bras and Oyvind Tafjord and Chris Wilhelm and Luca Soldaini and Noah A. Smith and Yizhong Wang and Pradeep Dasigi and Hannaneh Hajishirzi},
      year={2025},
      eprint={2411.15124},
      archivePrefix={arXiv},
      primaryClass={cs.CL},
      url={https://arxiv.org/abs/2411.15124}, 
}

@misc{hendrycks2021measuringmassivemultitasklanguage,
      title={Measuring Massive Multitask Language Understanding}, 
      author={Dan Hendrycks and Collin Burns and Steven Basart and Andy Zou and Mantas Mazeika and Dawn Song and Jacob Steinhardt},
      year={2021},
      eprint={2009.03300},
      archivePrefix={arXiv},
      primaryClass={cs.CY},
      url={https://arxiv.org/abs/2009.03300}, 
}

@misc{kumar2025mmauprochallengingcomprehensivebenchmark,
      title={{MMAU-Pro}: A Challenging and Comprehensive Benchmark for Holistic Evaluation of Audio General Intelligence},
      author={Sonal Kumar and Šimon Sedláček and Vaibhavi Lokegaonkar and Fernando López and Wenyi Yu and Nishit Anand and Hyeonggon Ryu and Lichang Chen and Maxim Plička and Miroslav Hlaváček and William Fineas Ellingwood and Sathvik Udupa and Siyuan Hou and Allison Ferner and Sara Barahona and Cecilia Bolaños and Satish Rahi and Laura Herrera-Alarcón and Satvik Dixit and Siddhi Patil and Soham Deshmukh and Lasha Koroshinadze and Yao Liu and Leibny Paola Garcia Perera and Eleni Zanou and Themos Stafylakis and Joon Son Chung and David Harwath and Chao Zhang and Dinesh Manocha and Alicia Lozano-Diez and Santosh Kesiraju and Sreyan Ghosh and Ramani Duraiswami},
      year={2025},
      eprint={2508.13992},
      archivePrefix={arXiv},
      primaryClass={eess.AS},
      url={https://arxiv.org/abs/2508.13992}, 
}

@misc{Weck_2026,
   title={{HumMusQA}: A Human-written Music Understanding {QA} Benchmark Dataset},
   author={Benno Weck and Pablo Puentes and Andrea Poltronieri and Satyajeet Prabhu and Dmitry Bogdanov},
   year={2026},
   eprint={2603.27877},
   archivePrefix={arXiv},
   primaryClass={cs.CL},
   doi={10.18653/v1/2026.nlp4musa-1.9},
   url={https://arxiv.org/abs/2603.27877},
}

@misc{carone2025musebenchmarkprobingmusic,
   title={The {MUSE} Benchmark: Probing Music Perception and Auditory Relational Reasoning in Audio {LLMs}},
   author={Brandon James Carone and Iran R. Roman and Pablo Ripollés},
   year={2025},
   eprint={2510.19055},
   archivePrefix={arXiv},
   primaryClass={cs.AI},
   url={https://arxiv.org/abs/2510.19055}, 
}

@misc{yuan2024chatmusicianunderstandinggeneratingmusic,
      title={{ChatMusician}: Understanding and Generating Music Intrinsically with {LLM}},
      author={Ruibin Yuan and Hanfeng Lin and Yi Wang and Zeyue Tian and Shangda Wu and Tianhao Shen and Ge Zhang and Yuhang Wu and Cong Liu and Ziya Zhou and Ziyang Ma and Liumeng Xue and Ziyu Wang and Qin Liu and Tianyu Zheng and Yizhi Li and Yinghao Ma and Yiming Liang and Xiaowei Chi and Ruibo Liu and Zili Wang and Pengfei Li and Jingcheng Wu and Chenghua Lin and Qifeng Liu and Tao Jiang and Wenhao Huang and Wenhu Chen and Emmanouil Benetos and Jie Fu and Gus Xia and Roger Dannenberg and Wei Xue and Shiyin Kang and Yike Guo},
      year={2024},
      eprint={2402.16153},
      archivePrefix={arXiv},
      primaryClass={cs.SD},
      url={https://arxiv.org/abs/2402.16153}, 
}

@misc{sourada2026musicicareabout,
      title={Music {I} Care About: Automated Multimodal Benchmarking of {LLM} Music Perception Skills on (Almost) Any Music},
      author={Tomáš Sourada and Katia Vendrame and Hajič, Jr., Jan},
      year={2026},
      eprint={2607.06015},
      archivePrefix={arXiv},
      primaryClass={cs.SD},
      url={https://arxiv.org/abs/2607.06015}, 
}

@misc{xu2025qwen3omnitechnicalreport,
      title={{Qwen3-Omni} Technical Report},
      author={Jin Xu and Zhifang Guo and Hangrui Hu and Yunfei Chu and Xiong Wang and Jinzheng He and Yuxuan Wang and Xian Shi and Ting He and Xinfa Zhu and Yuanjun Lv and Yongqi Wang and Dake Guo and He Wang and Linhan Ma and Pei Zhang and Xinyu Zhang and Hongkun Hao and Zishan Guo and Baosong Yang and Bin Zhang and Ziyang Ma and Xipin Wei and Shuai Bai and Keqin Chen and Xuejing Liu and Peng Wang and Mingkun Yang and Dayiheng Liu and Xingzhang Ren and Bo Zheng and Rui Men and Fan Zhou and Bowen Yu and Jianxin Yang and Le Yu and Jingren Zhou and Junyang Lin},
      year={2025},
      eprint={2509.17765},
      archivePrefix={arXiv},
      primaryClass={cs.CL},
      url={https://arxiv.org/abs/2509.17765}, 
}

@misc{gemmateam2026gemma4technicalreport,
      title={{Gemma 4} Technical Report},
      author={{Gemma Team}},
      year={2026},
      eprint={2607.02770},
      archivePrefix={arXiv},
      primaryClass={cs.CL},
      url={https://arxiv.org/abs/2607.02770}, 
}

@misc{goel2025audioflamingo3advancing,
      title={{Audio Flamingo 3}: Advancing Audio Intelligence with Fully Open Large Audio Language Models},
      author={Arushi Goel and Sreyan Ghosh and Jaehyeon Kim and Sonal Kumar and Zhifeng Kong and Lee, {Sang-gil} and Chao-Han Huck Yang and Ramani Duraiswami and Dinesh Manocha and Rafael Valle and Bryan Catanzaro},
      year={2025},
      eprint={2507.08128},
      archivePrefix={arXiv},
      primaryClass={cs.SD},
      url={https://arxiv.org/abs/2507.08128}, 
}

@misc{ghosh2026audioflamingonextnextgeneration,
      title={{Audio Flamingo Next}: Next-Generation Open Audio-Language Models for Speech, Sound, and Music},
      author={Sreyan Ghosh and Arushi Goel and Kaousheik Jayakumar and Lasha Koroshinadze and Nishit Anand and Zhifeng Kong and Siddharth Gururani and Lee, {Sang-gil} and Jaehyeon Kim and Aya Aljafari and Chao-Han Huck Yang and Sungwon Kim and Ramani Duraiswami and Dinesh Manocha and Mohammad Shoeybi and Bryan Catanzaro and Ming-Yu Liu and Wei Ping},
      year={2026},
      eprint={2604.10905},
      archivePrefix={arXiv},
      primaryClass={cs.SD},
      url={https://arxiv.org/abs/2604.10905}, 
}

@misc{nvidia2026nemotron3nanoomni,
      title={{Nemotron 3 Nano Omni}: Efficient and Open Multimodal Intelligence},
      author={{NVIDIA}},
      year={2026},
      eprint={2604.24954},
      archivePrefix={arXiv},
      primaryClass={cs.LG},
      url={https://arxiv.org/abs/2604.24954}, 
}

@misc{kong2026unifiedaudiointelligenceregressing,
      title={Unified Audio Intelligence Without Regressing on Text Intelligence}, 
      author={Zhifeng Kong and Lee, {Sang-gil} and Jaehyeon Kim and Boxin Wang and Zihan Liu and Sungwon Kim and Yang Chen and Arushi Goel and Rajarshi Roy and Wenliang Dai and Zhuolin Yang and Yangyi Chen and Dongfu Jiang and Sreyan Ghosh and Tuomas Rintamaki and Andrew Tao and Jonathan Raiman and Mohammad Shoeybi and Bryan Catanzaro and Wei Ping},
      year={2026},
      eprint={2607.05196},
      archivePrefix={arXiv},
      primaryClass={cs.CL},
      url={https://arxiv.org/abs/2607.05196}, 
}

@misc{yang2026midillamainstructionfollowingmultimodalllm,
      title={{MIDI-LLaMA}: An Instruction-Following Multimodal {LLM} for Symbolic Music Understanding},
      author={Meng Yang and Jon McCormack and Maria Teresa Llano and Wanchao Su and Chao Lei},
      year={2026},
      eprint={2601.21740},
      archivePrefix={arXiv},
      primaryClass={cs.MM},
      url={https://arxiv.org/abs/2601.21740}, 
}

@misc{kong2022giantmidipianolargescalemididataset,
      title={{GiantMIDI-Piano}: A large-scale {MIDI} dataset for classical piano music},
      author={Qiuqiang Kong and Bochen Li and Jitong Chen and Yuxuan Wang},
      year={2022},
      eprint={2010.07061},
      archivePrefix={arXiv},
      primaryClass={cs.IR},
      url={https://arxiv.org/abs/2010.07061}, 
}

@article{benetos2018automatic,
  title={Automatic music transcription: An overview},
  author={Benetos, Emmanouil and Dixon, Simon and Duan, Zhiyao and Ewert, Sebastian},
  journal={IEEE Signal Processing Magazine},
  volume={36},
  number={1},
  pages={20--30},
  year={2019},
  publisher={IEEE}
}

@misc{defossez2022hybridspectrogramwaveformsource,
      title={Hybrid Spectrogram and Waveform Source Separation}, 
      author={Alexandre Défossez},
      year={2022},
      eprint={2111.03600},
      archivePrefix={arXiv},
      primaryClass={eess.AS},
      url={https://arxiv.org/abs/2111.03600}, 
}

@misc{ma2025mmarchallengingbenchmarkdeep,
      title={{MMAR}: A Challenging Benchmark for Deep Reasoning in Speech, Audio, Music, and Their Mix},
      author={Ziyang Ma and Yinghao Ma and Yanqiao Zhu and Chen Yang and Yi-Wen Chao and Ruiyang Xu and Wenxi Chen and Yuanzhe Chen and Zhuo Chen and Jian Cong and Kai Li and Keliang Li and Siyou Li and Xinfeng Li and Xiquan Li and Zheng Lian and Yuzhe Liang and Minghao Liu and Zhikang Niu and Tianrui Wang and Yuping Wang and Yuxuan Wang and Yihao Wu and Guanrou Yang and Jianwei Yu and Ruibin Yuan and Zhisheng Zheng and Ziya Zhou and Haina Zhu and Wei Xue and Emmanouil Benetos and Kai Yu and Eng-Siong Chng and Xie Chen},
      year={2025},
      eprint={2505.13032},
      archivePrefix={arXiv},
      primaryClass={cs.SD},
      url={https://arxiv.org/abs/2505.13032}, 
}

@misc{solovyev2024benchmarksleaderboardssounddemixing,
      title={Benchmarks and leaderboards for sound demixing tasks}, 
      author={Roman Solovyev and Alexander Stempkovskiy and Tatiana Habruseva},
      year={2024},
      eprint={2305.07489},
      archivePrefix={arXiv},
      primaryClass={cs.SD},
      url={https://arxiv.org/abs/2305.07489}, 
}

@misc{sekoyan2025canary1bv2parakeettdt06bv3efficient,
      title={{Canary-1B-v2} \& {Parakeet-TDT-0.6B-v3}: Efficient and High-Performance Models for Multilingual {ASR} and {AST}},
      author={Monica Sekoyan and Nithin Rao Koluguri and Nune Tadevosyan and Piotr Zelasko and Travis Bartley and Nikolay Karpov and Jagadeesh Balam and Boris Ginsburg},
      year={2025},
      eprint={2509.14128},
      archivePrefix={arXiv},
      primaryClass={cs.CL},
      url={https://arxiv.org/abs/2509.14128}, 
}

@misc{bradshaw2025scalingselfsupervisedrepresentationlearning,
      title={Scaling Self-Supervised Representation Learning for Symbolic Piano Performance}, 
      author={Louis Bradshaw and Honglu Fan and Alexander Spangher and Stella Biderman and Simon Colton},
      year={2025},
      eprint={2506.23869},
      archivePrefix={arXiv},
      primaryClass={cs.SD},
      url={https://arxiv.org/abs/2506.23869}, 
}
\bibliographystyle{iclr2027_conference}

\clearpage
\appendix
\crefalias{section}{appendix}
\crefalias{subsection}{appendix}
\section{Video Discovery and Selection}
\label{appendix:video-selection}

\subsection{Candidate discovery}
\label{appendix:candidate-discovery}

We score video titles and the first 512 characters of their descriptions for
likely piano content, beginning with manually selected seeds. The first prompt
assigns scores from 0 to 6. Scores 0, 1, and 2 indicate no evidence of piano,
incidental piano, and piano-adjacent content, respectively. Score 3 indicates
piano-related content with uncertain solo-piano audio; scores 4 and 5 indicate
likely and very likely contiguous piano-only audio. Score 6 additionally
requires evidence of uncommon repertoire, improvisation, or a distinctive
arrangement. Videos scoring at least 5 seed searches through related videos;
videos scoring 3 are evaluated by the speech-piano prompt below.

We expand a channel when at least five of its videos pass the speech-piano
metadata filter, or when more than 25\% of its scored videos have piano-content
scores of at least 5. Videos found through either route enter the same scoring
process.

\subsection{Speech--piano selection}
\label{appendix:speech-piano-selection}
The second prompt scores whether a video contains spoken commentary and related
solo-piano examples. It assigns 0 to unlikely matches, 4 to possible matches,
5 to likely matches, and 6 to clear matches. Scores of at least 4 pass the
metadata filter. The prompt is reproduced in full below.

\begingroup
\scriptsize
\begin{verbatim}
Analyze the YouTube video title and description.

Task:
Score how likely the video is to contain useful paired spoken language and
solo-piano audio.

The intended dataset uses timed captions together with nearby piano examples.
Speech and piano may alternate or overlap. Useful videos should likely contain:

1. Meaningful spoken instruction, explanation, analysis, or commentary
2. Audible solo-piano examples in the same video
3. A meaningful relationship between the speech and piano examples

The video was selected from piano-related metadata, but piano playing is not
guaranteed. Judge only from the supplied title and description.

Good candidates include spoken piano tutorials or lessons, harmony/theory
breakdowns, masterclasses, technique explanations, composition or performance
analysis at the piano, and narrated piano or virtual-piano demonstrations.

Use only these scores:

6 = Clear high-yield match.
The metadata explicitly indicates substantial spoken explanation together with
related solo-piano demonstrations, passages, or examples.

5 = Likely useful candidate.
The metadata suggests meaningful speech and related solo-piano examples in the
same video, but does not give enough concrete detail for 6.

4 = Possible candidate.
The metadata is sparse or ambiguous, but related speech and solo-piano examples
are plausible. This includes ordinary piano tutorials or lessons unless the
metadata clearly indicates a non-verbal format.

0 = Unlikely or wrong format.
Use 0 for performance-only videos, speech without related piano examples,
unrelated background piano, vocals, ensemble, or clearly non-verbal piano
content.

Do not infer spoken narration merely because the written description explains
the music. Interviews, reactions, reviews, and discussions score 5 or 4 only
when the metadata also suggests related solo-piano examples in the same video.

If both title and description are missing or blank, use 0.

Return only JSON with one of the scores 6, 5, 4, or 0. Do not use a markdown
fence:
{"score": N}
\end{verbatim}
\endgroup

\subsection{Measuring interleaving}
\label{appendix:interleaving}

During dataset construction, recordings were selected if they had piano and
speech content independently. To quantify the degree of interleaving, we split
speech and MIDI independently at pauses longer than 2 seconds, then add
candidate boundaries within longer passages using a 5-second window. We sort
these candidate chunks by start time and count adjacent speech-MIDI switches.
For recordings with both modalities, $N>2$ candidates, and $S$ switches, the
score is $100(S-1)/(N-2)$. A recording with both modalities but only one
speech--MIDI switch scores zero, while switching at every boundary scores 100.
Single-modality recordings receive a score of zero.
\cref{fig:interleaving-distribution} shows the distribution of these scores
across the corpus.

\section{\benchmark{} Construction Details}
\label{appendix:benchmark-construction}

\subsection{Candidate Timelines and Video Ranking}
\label{appendix:benchmark-timelines}

The following timeline records begin in the first minute of the test-split
video that produced released question 350. Speech about a tritone substitution
alternates with five piano regions, including the two-note example beginning at
53.07\,s. We reconstructed this excerpt from the released YouTube captions and
full-length MIDI.
Each JSON object is one record.

\begin{Verbatim}[fontsize=\scriptsize,breaklines=true,breakanywhere=true]
{"type": "speech", "start_s": 0.12, "end_s": 4.92, "text": "this tune uses one of the biggest chord substitutions in"}
{"type": "speech", "start_s": 4.92, "end_s": 5.71, "text": "jazz"}
{"type": "piano", "start_s": 5.84, "end_s": 18.31, "number_of_notes_in_segment": 109}
{"type": "speech", "start_s": 18.08, "end_s": 22.97, "text": "that satin doll the Duke Ellington hit written by Billy strayhorn and the"}
{"type": "speech", "start_s": 22.98, "end_s": 26.48, "text": "classic substitution that it uses is called a tritone"}
{"type": "speech", "start_s": 26.48, "end_s": 31.37, "text": "substitution G7 has these two really really important"}
{"type": "piano", "start_s": 27.63, "end_s": 29.06, "number_of_notes_in_segment": 4}
{"type": "piano", "start_s": 30.23, "end_s": 31.73, "number_of_notes_in_segment": 10}
{"type": "speech", "start_s": 31.38, "end_s": 36.12, "text": "notes that are in them another chord that has those two really really important notes"}
{"type": "speech", "start_s": 36.12, "end_s": 40.92, "text": "is d flat seven it also has those"}
{"type": "piano", "start_s": 38.2, "end_s": 40.25, "number_of_notes_in_segment": 16}
{"type": "speech", "start_s": 40.92, "end_s": 45.36, "text": "notes so if both those chords have the important notes well you can just replace"}
{"type": "speech", "start_s": 45.36, "end_s": 50.28, "text": "one for another and that's what happens in satin doll so that's the substitution part well why is"}
{"type": "speech", "start_s": 50.28, "end_s": 55.199, "text": "it called a tritone substitution well G and D"}
{"type": "piano", "start_s": 53.07, "end_s": 58.09, "number_of_notes_in_segment": 2}
{"type": "speech", "start_s": 55.199, "end_s": 60.059, "text": "flat are a tritone apart so if satin all didn't have a tritone"}
\end{Verbatim}

Gemma 4 26B-A4B assigns each video an interest score from 1 to 6 based on its
timeline. Higher-scoring videos are processed first for question generation.
The complete ranking prompt is shown below. Angle-bracketed text denotes the
runtime timeline substitution; all other text is reproduced verbatim.

\begin{Verbatim}[fontsize=\scriptsize,breaklines=true,breakanywhere=true]
Judge the strongest part of this timeline as a source for an audio-only piano
question. One strong part is enough; ignore video length and opportunity count.

The goal is a benchmark of musical understanding. The answer to an interesting
question should preserve a specific insight worth learning about the logic or
significance of the music, not merely the mechanics of producing it or an
obvious feature of the sound.

6 = Excellent material for an exceptionally revealing advanced benchmark question.
5 = Clear evidence of an advanced, substantive musical insight.
4 = Possibly advanced, with important uncertainty.
3 = Meaningful but intermediate or borderline musical understanding.
2 = Only routine, introductory, surface-level, or instrumental-mechanics material.
1 = No plausible meaningful material.

If the strongest material is only an instrumental technique or practice
pattern, it belongs at 2 or 3 even when it is difficult, systematic, or
organized using music theory.

Give 6 when the speech makes a clear, advanced insight available through the
associated piano music: the resulting question and answer could retain some of
the explanatory value of the source. The insight need not be novel to experts.
Judge substance, not impressive vocabulary. Assume speech and piano are
generally well paired. Do not propose questions.

Return only JSON:
{"score": N}

TIMELINE:
<timeline JSON objects, one per line>
\end{Verbatim}

\subsection{Question Generation}
\label{appendix:question-generation}

GPT 5.6 Sol receives the timeline and proposes up to five questions per video,
each with an integer-second clip interval no longer than 20\,s, a fixed
correct answer, and an evidence summary. It does not receive the piano audio
or detailed MIDI notes. The speech text guides it toward candidate passages,
but the eventual respondent receives only the isolated piano passage, question,
and choices. The prompt below presents the question-generation instructions
in API form: instructions about agent file and tool access are omitted, while
the content criteria are retained. The runtime timeline substitution is marked
in angle brackets.

\begin{Verbatim}[fontsize=\scriptsize,breaklines=true,breakanywhere=true]
Create benchmark-style questions about isolated piano audio.

All source information is included in this prompt. Work only from that
information.

Input:

TIMELINE: A chronological sequence of JSON objects, one per line, ordered by
each region's onset time.

- A speech region contains `type: "speech"`, `start_s`, `end_s`, and `text`.
- A piano region contains `type: "piano"`, `start_s`, `end_s`, and
  `number_of_notes_in_segment`.

Read the timeline in order and use the temporal relationship between speech and
nearby piano regions to locate piano passages that may support clear audio-only
questions. Speech is hidden metadata and will not be available to the listener.
Piano regions describe the source-separated, piano-only audio that will be
available to the listener.

Regions may overlap even though they are ordered by onset. Use their explicit
start and end times rather than assuming that one region ends when the next
begins. Continuous speech may be divided into consecutive timestamped regions
without a silence between them.

The detected note count confirms piano activity and may provide a rough clue
about musical density. It does not reveal the pitches, chords, rhythm,
articulation, technique, or musical identity of the passage.

For each QA pair, `audio_timestamp` is used to extract a standalone clip from
the source-separated piano audio. The listener receives only that clip, the
question, and the choices. Playback starts at 0 and the source timeline is not
available. Source timestamps are extraction metadata only and must not appear
in the question or choices. Refer to positions within the clip using terms such
as "at the beginning," "later," or "at the end."

Task:

Aim to generate 2 to 5 interesting and reliable questions about musical
content by turning concrete spoken observations about nearby piano passages
into distinctions that should be audible in a short isolated clip. Use
independent local events or passages where possible. Return fewer than two,
including zero, only when the supplied timeline genuinely lacks enough
concrete, speech-tied piano events to support them.

For each retained question, establish exactly one fixed correct answer, the
shortest practical integer-second clip, a concise timeline-supported evidence
summary, evidence provenance, and uncertainty. Include only supplied evidence;
do not imply that you heard the audio or invent musical details. Do not
generate, propose, or consider any incorrect answer choices yourself. They are
generated in a later stage.

Audio-grounding rules:

- Design each question so that a listener should be able to answer it from the
  selected piano clip without transcript, speech, visuals, or outside context.
- Do not refer to speakers, instructors, narration, transcripts, videos, lessons,
  or visual content.
- Do not ask what was said, intended, named, requested, or explained.
- Use speech to infer candidate questions about nearby piano passages. Because
  you cannot hear the audio, treat spoken descriptions as evidence rather than
  guaranteed truth.
- Favor concrete musical claims that speech ties to a specific passage and that
  are likely to be clearly audible in the isolated piano clip.
- The extracted clip must be no longer than 20 seconds and must contain all
  audio needed to answer the question.
- Choose the shortest practical audio range. Prefer a single local event or
  compact contrast; do not span distant passages or include long intervening
  audio.
- Do not infer specific musical details from note count alone.
- A comparison, change, or before-and-after question must include all passages
  needed for the comparison within the selected clip.
- Reject claims that require speech, notation, visuals, or facts not established
  by the audio, as well as musical details that would be too subtle to verify
  reliably. Recognition of conventional musical characteristics and categories
  from sound counts as audio understanding, not outside knowledge.

Question and correct-answer rules:

- Each question must have exactly one correct answer that is clearly supported
  by the audio.
- Avoid yes/no questions, subjective judgments, trivia, generic activity
  labels, and questions requiring perfect pitch unless the clip supplies an
  audible reference pitch.
- Consider the full range of musically meaningful information recoverable from
  the clip, from local audible details and relationships to broader
  characterization of the passage. Do not default to instructional details or
  before-and-after demonstrations merely because they are easier to locate.
- Fix the question and correct answer before distractor generation. Do not
  revise either one based on the distractor candidates.

Return JSON only, using this schema:

[
  {
    "id": 0,
    "audio_timestamp": [start_second, end_second],
    "question": "Question text",
    "correct_answer": "Correct answer",
    "evidence": "Concise description supported by the timeline",
    "provenance": ["Relevant supplied timeline observation"],
    "uncertainty": "low"
  }
]

IDs must be consecutive from zero. Both `audio_timestamp` values must be
nonnegative integer seconds from the start of the source, the end must be
greater than the start and no more than 20 seconds later. `provenance` must be a
nonempty list of concise strings, and `uncertainty` must be `low`, `medium`, or
`high`.

TIMELINE:
<timeline JSON objects, one per line>
\end{Verbatim}

\subsection{Distractor Generation and Scoring}
\label{appendix:distractor-generation}

For each fixed question and answer, GPT 5.6 Terra proposes ten candidate
distractors using the evidence summary from question generation. It removes
choices it judges redundant, ambiguous, potentially correct, or irrelevant
before scoring. The model has no access to the piano audio. The prompt below
presents these instructions in API form, with runtime question data marked in
angle brackets. The implementation supplied this data and separately scored
the returned distractors.

\begin{Verbatim}[fontsize=\scriptsize,breaklines=true,breakanywhere=true]
Generate distractors for one fixed piano-audio question. Use only the complete
question data supplied below. You do not have the audio.

Generate `candidate_count` plausible incorrect answers. Each candidate must be
clearly wrong given the supplied textual audio evidence, while remaining
plausible to a model that sees only the question and answer choices. Match the
correct answer's musical dimension, grammar, specificity, length, and
terminology; prefer small counterfactual changes over conspicuous foils.

Before returning, remove candidates that are duplicated, semantically redundant,
ambiguous, potentially correct, off-axis, malformed, or unsupported by the
evidence. Keep at least three materially distinct candidates.

Return JSON only, using this shape:

{
  "candidates": ["Candidate 1", "Candidate 2", "Candidate 3"]
}

COMPLETE QUESTION DATA

{
  "question_id": <question_id>,
  "audio_timestamp": [<start_second>, <end_second>],
  "question": "<question>",
  "correct_answer": "<correct_answer>",
  "evidence": "<evidence>",
  "provenance": <provenance JSON array>,
  "uncertainty": "<uncertainty>",
  "candidate_count": 10
}
\end{Verbatim}

For each remaining candidate distractor, the scoring tool constructs three four-choice
sets containing the correct answer, that distractor, and two other sampled
distractors. It shuffles the choices and asks Gemma 4 12B to answer using only
the question and choices. The distractor's adversarial score is the mean probability Gemma
assigns to the distractor's choice letter across the three sets. This is a
ranking of how tempting the incorrect answer appears without music; it is not
a probability that the answer is musically correct. The text-only scoring
prompt follows. The \texttt{\{question\}} placeholder receives the question
text, and \texttt{\{choices\}} receives the four labelled answer choices.

\begin{Verbatim}[fontsize=\scriptsize,breaklines=true,breakanywhere=true]
Answer this music multiple-choice question using only its text. Choose the
option that appears most likely to be correct.

QUESTION:
{question}

CHOICES:
{choices}

Return only the single letter A, B, C, or D.
\end{Verbatim}

\subsection{A Released Benchmark Example}
\label{appendix:benchmark-example}

After listening to the separated piano audio and the MIDI rendering, an author
retained questions whose answer was unambiguous under both representations and
rejected unsuitable distractors. The highest-scoring three remaining
distractors were selected, and the four choices were shuffled once for the
released benchmark. The following is the complete plain-text MIDI input sent
to the evaluated model for question 1037, copied from the released
\benchmark{} dataset. Long lines wrap only for display. The clip lasts six
seconds.

\begin{Verbatim}[fontsize=\scriptsize,breaklines=true,breakanywhere=true]
Answer the multiple-choice question about the isolated piano performance.
Use only the compact piano note table below as evidence. Choose exactly one
option.

Table interpretation:
- `unit` is the duration of one timing unit; `advance` is the duration of one
  time window.
- Each `midi_pitch,velocity,onset,duration;` record describes one note; MIDI
  pitch 60 is middle C.
- `onset` is measured in timing units from the start of the current window.
- `duration` is measured in timing units and may extend beyond that window.
- Each `advance;` moves the current window forward by the header's `advance`
  duration.

PIANO NOTE TABLE:
<piano>unit=10ms;advance=5s;notes(midi_pitch,velocity,onset,duration):41,50,0,53;45,60,50,47;48,60,94,48;50,40,140,47;51,70,185,46;50,60,231,46;48,60,276,41;45,60,316,45;41,60,360,96;</piano>

QUESTION:
What kind of bass accompaniment pattern is played?

CHOICES:
A. An Alberti bass pattern.
B. A walking bass line.
C. A root-and-fifth bass pattern.
D. A pedal-point bass line.

Return only a JSON object with a single "choice" field whose value is one of
"A", "B", "C", or "D".
\end{Verbatim}

\noindent The correct choice is B, a walking bass line.

\section{Benchmark Evaluation Details}

\subsection{Model Inference Settings}
\label{appendix:model-inference}

\Cref{tab:model-inference-settings} reports the requested inference settings
for each configuration in \cref{tab:benchmark-results}. We asked all models to
return a JSON object containing one answer letter. This specifies the final
response format but does not disable a model's reasoning process. A dash in
the sampling columns means that the parameter was omitted from the request.
In the thinking column, it means that thinking was disabled or no separate
thinking setting was available. Repeated model identifiers
correspond to evaluations with different thinking settings. For Qwen 3 Omni
and Gemma 4 12B, 1024 denotes the requested thinking-token budget.

\begin{table*}[t]
    \centering
    \renewcommand{\arraystretch}{1.2}
    \setlength{\tabcolsep}{3pt}
    \footnotesize
    \begin{tabular*}{\textwidth}{@{\extracolsep{\fill}}>{\raggedright\arraybackslash}p{0.28\textwidth}>{\scriptsize\raggedright\arraybackslash}p{0.43\textwidth}ccp{0.10\textwidth}@{}}
        \toprule
        \textbf{Model} & \textbf{Inference route} & \textbf{Temp.} & \textbf{Top-$p$} & \textbf{Thinking} \\
        \midrule
        \multicolumn{5}{@{}l}{\textit{Self-hosted}} \\
        \addlinespace[0.3ex]
        Qwen 3 Omni (thinking) & \nolinkurl{huggingface:Qwen/Qwen3-Omni-30B-A3B-Thinking} & 0.6 & 0.95 & 1024 \\
        Qwen 3 Omni (instruct) & \nolinkurl{huggingface:Qwen/Qwen3-Omni-30B-A3B-Instruct} & 0 & 1 & -- \\
        Gemma 4 12B (thinking) & \nolinkurl{huggingface:google/gemma-4-12B-it} & 1 & 0.95 & 1024 \\
        Gemma 4 12B (non-thinking) & \nolinkurl{huggingface:google/gemma-4-12B-it} & 0 & 1 & -- \\
        Nemotron Audex 30B-A3B & \nolinkurl{huggingface:nvidia/Nemotron-Labs-Audex-30B-A3B} & 0 & 1 & -- \\
        Audio Flamingo 3 & \nolinkurl{huggingface:nvidia/audio-flamingo-3-hf} & 0 & 1 & -- \\
        Music Flamingo & \nolinkurl{huggingface:nvidia/music-flamingo-hf} & 0 & 1 & -- \\
        Qwen 3.5 9B (thinking) & \nolinkurl{huggingface:Qwen/Qwen3.5-9B} & 1 & 0.95 & 16000 \\
        Qwen 3.5 9B (non-thinking) & \nolinkurl{huggingface:Qwen/Qwen3.5-9B} & 0 & 1 & -- \\
        \midrule
        \multicolumn{5}{@{}l}{\textit{Provider-hosted}} \\
        \addlinespace[0.3ex]
        Nemotron 3 Nano Omni & \nolinkurl{openrouter:nvidia/nemotron-3-nano-omni-30b-a3b-reasoning:free} & 0.6 & 0.95 & Enabled \\
        Inkling$_{\text{small}}$ (medium) & \nolinkurl{openrouter:thinkingmachines/inkling-small} & 1 & 1 & Medium \\
        Inkling (medium) & \nolinkurl{openrouter:thinkingmachines/inkling} & 1 & 1 & Medium \\
        Gemini 3.1 Pro (medium) & \nolinkurl{openrouter:google/gemini-3.1-pro-preview} & -- & -- & Medium \\
        Qwen 3.5 Omni Plus & \nolinkurl{alibaba:qwen3.5-omni-plus-2026-03-15} & -- & -- & -- \\
        Muse Spark 1.2 (medium) & \nolinkurl{openrouter:meta/muse-spark-1.2} & 0 & 1 & Medium \\
        Gemini 3.7 Flash (medium) & \nolinkurl{openrouter:google/gemini-3.7-flash} & -- & -- & Medium \\
        Gemini 3.8 Flash (medium) & \nolinkurl{openrouter:google/gemini-3.8-flash} & -- & -- & Medium \\
        GPT 5.6 Terra (high) & \nolinkurl{openai:gpt-5.6-terra} & -- & -- & High \\
        GPT 5.6 Sol (none) & \nolinkurl{openai:gpt-5.6-sol} & -- & -- & -- \\
        GPT 5.6 Sol (medium) & \nolinkurl{openai:gpt-5.6-sol} & -- & -- & Medium \\
        GPT 5.6 Sol (high) & \nolinkurl{openai:gpt-5.6-sol} & -- & -- & High \\
        GPT 5.6 Sol (xhigh) & \nolinkurl{openai:gpt-5.6-sol} & -- & -- & Xhigh \\
        GPT 6 Astra (high) & \nolinkurl{openai:gpt-6-astra} & -- & -- & High \\
        \bottomrule
    \end{tabular*}
    \caption{Requested temperature, top-$p$, and thinking settings for the models evaluated on \benchmark{}. Model labels abbreviate those in \cref{tab:benchmark-results} where the identity remains clear.}
    \label{tab:model-inference-settings}
\end{table*}

\raggedbottom
\subsection{Evaluation Prompts}
\label{appendix:evaluation-prompts}

\noindent\begin{minipage}{\linewidth}
\setlength{\parskip}{.5pc}
\paragraph{No-music control.} The model receives the question and four answer
choices without music. The prompt is shown below; \texttt{\{question\}} and
\texttt{\{choices\}} are replaced at evaluation time.
\par\medskip

\begin{Verbatim}[fontsize=\scriptsize,breaklines=true,breakanywhere=true]
Answer the multiple-choice question about an isolated piano performance.
No representation of the performance is provided; answer using only the
question and answer choices. Choose exactly one option.

QUESTION:
{question}

CHOICES:
{choices}

Return only a JSON object with a single "choice" field whose value is one of
"A", "B", "C", or "D".
\end{Verbatim}
\end{minipage}
\par\medskip

\noindent\begin{minipage}{\linewidth}
\setlength{\parskip}{.5pc}
\paragraph{MIDI-text.} The model receives the same question and choices with
a transcript of the piano passage encoded as MIDI-text. The additional
\texttt{\{midi\_document\}} placeholder is replaced with the note table.
\par\medskip

\begin{Verbatim}[fontsize=\scriptsize,breaklines=true,breakanywhere=true]
Answer the multiple-choice question about the isolated piano performance.
Use only the compact piano note table below as evidence. Choose exactly one
option.

Table interpretation:
- `unit` is the duration of one timing unit; `advance` is the duration of one
  time window.
- Each `midi_pitch,velocity,onset,duration;` record describes one note; MIDI
  pitch 60 is middle C.
- `onset` is measured in timing units from the start of the current window.
- `duration` is measured in timing units and may extend beyond that window.
- Each `advance;` moves the current window forward by the header's `advance`
  duration.

PIANO NOTE TABLE:
{midi_document}

QUESTION:
{question}

CHOICES:
{choices}

Return only a JSON object with a single "choice" field whose value is one of
"A", "B", "C", or "D".
\end{Verbatim}
\end{minipage}
\par\medskip

\noindent\begin{minipage}{\linewidth}
\setlength{\parskip}{.5pc}
\paragraph{Audio.} The model receives the same question and choices with an
audio attachment. We use this prompt for both the source-separated piano stem
and the Pianoteq rendering.
\par\medskip

\begin{Verbatim}[fontsize=\scriptsize,breaklines=true,breakanywhere=true]
Answer the multiple-choice question about the isolated piano performance.
Use only the attached audio as evidence. Choose exactly one option.

QUESTION:
{question}

CHOICES:
{choices}

Return only a JSON object with a single "choice" field whose value is one of
"A", "B", "C", or "D".
\end{Verbatim}
\end{minipage}
\par\medskip

\clearpage
\subsection{GPT 6 Astra Error Analysis}
\label{appendix:astra-errors}

\Cref{tab:astra-errors} lists the 17 questions GPT 6 Astra answered
incorrectly with MIDI-text input. The \#Notes column gives the number of notes
in the supplied transcription. The Error column indicates whether there are
transcription errors that affect the question.

\begin{center}
  \begin{minipage}{\textwidth}
    \centering
    \renewcommand{\arraystretch}{1.15}
    \setlength{\tabcolsep}{4pt}
    \footnotesize
    \begin{tabular*}{\textwidth}{@{\extracolsep{\fill}}r>{\raggedright\arraybackslash}p{0.72\textwidth}r c@{}}
        \toprule
        \textbf{ID} & \textbf{Question} & \textbf{\#Notes} & \textbf{Error} \\
        \midrule
        36 & What rhythmic style characterizes the left-hand accompaniment? & 255 & No  \\ 
        99 & What articulation best describes the notes in the short repeated figure? & 59 & No  \\
        128 & What is the relationship between the low repeated notes and the melody? & 113 & No  \\
        138 & What recurring pulse pattern supports the melody? & 44 & No  \\
        157 & Where does the harmony move later in the clip relative to the opening chord? & 8 & Yes  \\
        182 & How does the chord quality change from the first chord to the next? & 3 & Yes  \\
        393 & How many times does the brief four-note figure recur in the clip? & 3 & Yes  \\
        564 & What chord quality is heard in the passage? & 9 & No  \\
        614 & What change in chord quality redirects the progression toward a new key? & 24 & No \\
        665 & Which tonal mode characterizes the passage? & 108 & No  \\
        676 & What overall character is established at the beginning of this passage? & 205 & No  \\
        685 & How many notes make up the chord heard in this clip? & 2 & Yes  \\
        730 & What is the overall direction of the chord progression in this clip? & 36 & No  \\
        732 & What harmonic-quality change occurs later in the clip? & 57 & No  \\
        733 & What kind of meter and rhythmic feel characterizes this passage? & 47 & No  \\
        758 & What type of meter or dance-like accompaniment underpins this passage? & 91 & No  \\
        1031 & How many contrapuntal voices are sounding in the texture? & 72 & No  \\
        \bottomrule
    \end{tabular*}
    \captionof{table}{Questions answered incorrectly by GPT 6 Astra with MIDI-text input. The 17 errors correspond to 93.2\% accuracy on 250 questions. Question 564 has an incorrect human-selected reference answer rather than a transcription error: the notes C--E$\flat$--G--B form a minor--major seventh chord, not a minor seventh chord.}
    \label{tab:astra-errors}
  \end{minipage}
\end{center}

\section{Mid-Training Document Construction}
\label{appendix:mid-training-documents}

To produce interleaved speech and MIDI-text documents, we: (1)
Independently split speech and MIDI streams at gaps longer than 2 seconds.
(2) Within each resulting span, add a candidate boundary at an event
onset once at least 5 seconds have elapsed since the previous boundary,
provided at least 5 seconds of material remains. (3) Combine speech
and MIDI chunks and sort them by start time. (4) Merge consecutive
chunks of the same modality.

This preserves continuous single-modality passages while producing finer-grained
interleaving where speech and piano overlap. We then produce the training
documents by concatenating newline-separated chunks, wrapping MIDI-text in
\texttt{<piano>} XML tags. We form the separated documents by splitting each
final interleaved document into its speech and MIDI lines, without additional
segmentation or merging. The speech-only condition uses only the resulting
speech documents. \Cref{fig:mid-training-pipeline} visualizes this process.
We release the exact mid-training documents used in these experiments for
reproducibility.
Finally, during training we shuffle documents, tokenize them with the model's
tokenizer, append EOS, and pack the resulting stream into 4,096-token training
sequences.

\end{document}